\documentclass[amssymb,preprint, floats,amsmath,superscriptaddress,prb,aps]{revtex4-2}
\usepackage{setspace}
\setcitestyle{numbers,round}
\usepackage{graphicx}
\usepackage{textcomp}
\usepackage{float}
\usepackage[dvipsnames]{xcolor}
\usepackage[normalem]{ulem}
\usepackage{upgreek}
\usepackage[english]{babel}
\usepackage[utf8]{inputenc}
\usepackage{tikz}
\usepackage{braket}
\usepackage{cleveref}
\usepackage{amssymb}
\usepackage[colorinlistoftodos, color=green!40, prependcaption]{todonotes}
\usepackage{setspace}

\usepackage{amsthm}
\usepackage{mathtools}
\usepackage{physics}
\usepackage{xcolor}
\usepackage{graphicx}
\usepackage[left=23mm,right=13mm,top=35mm,columnsep=15pt]{geometry} 
\usepackage{adjustbox}
\usepackage{placeins}
\usepackage[T1]{fontenc}
\usepackage{csquotes}



\def \UT{MESA+ Institute, University of Twente, Enschede, the Netherlands.}
\def \UvA{Van der Waals–Zeeman Institute, IoP, University of Amsterdam, Amsterdam, the Netherlands.}

\begin{document}

\title{Coexisting topological hinges and 1D Rashba states in Bi$_{0.97}$Sb$_{0.03}$ revealed by the Josephson effect}

\author{Biplab Bhattacharyya}\affiliation{\UT}
\author{Stijn R. de Wit}\affiliation{\UT}
\author{Zhen Wu}\affiliation{\UT}
\author{Yingkai Huang}\affiliation{\UvA}
\author{Mark S. Golden}\affiliation{\UvA}
\author{Alexander Brinkman}\affiliation{\UT}
\author{Chuan Li$^*$}\affiliation{\UT}
\date{\today} 

\def\BB{\textcolor{blue}}
\def\CL{\textcolor{RoyalBlue}}
\def\AB{\textcolor{magenta}}
\def\SW{\textcolor{red}}

\def\BS{Bi$_{0.97}$Sb$_{0.03}$}

    \begin{abstract}
         Second-order topological insulating (SOTI) states in three-dimensional materials are helical one-dimensional hinge states. Inducing superconductivity in these states leads to gapless bound states, characterized by the 4$\pi$-periodic current-phase relation. Here, we provide evidence of the topologically protected hinge states in Dirac semimetal Bi$_{0.97}$Sb$_{0.03}$ nanoflakes by an unconventional interference pattern in a magnetic field, and the 4$\pi$-periodic supercurrent carried by these states via the suppressed first and third Shapiro steps. Tight-binding simulations confirm the presence of multiple hinge modes, supporting our interpretation of Bi$_{0.97}$Sb$_{0.03}$ as a prototypical designable SOTI platform. Quantum confinement effect is identified by a quasi-one-dimensional bulk transport, and the confined Rashba states are responsible for the broadened hinge states.
    \end{abstract}
 
\maketitle


\noindent {\small {*}Corresponding author: C.~Li, Email: \href{mailto:chuan.li@utwente.nl}{\textcolor{blue}{\underline{chuan.li@utwente.nl}}}}

\section{Introduction}
		\label{sect:intro}  
		\FloatBarrier

		\noindent Inducing superconductivity into topological materials in hybrid devices holds great interest for building topological quantum computing~\cite{Fu3DTI2007, Hasan2010RMP}. The topological protection of the surface states is manifested by the spin-momentum locking and the suppression of back scattering. When superconductivity is induced in the helical states via the Josephson effect, parity-protected gapless bound states will form, namely the Majorana bound states. Although such topological states can be found in different dimensions, their one-dimensional form provides the most significant physics due to its restricted spin degree of freedom. The 1D helical states can be realized as the edge state of 2D topological insulator (TI) system (namely the quantum spin Hall states) when the time-reversal symmetry and inversion symmetry are preserved~\cite{Bernevig2006, Konig2007} or the recently-discovered hinge states in 3D second-order TIs (SOTIs) which are protected by crystalline symmetries in three dimensions~\cite{Po2017, Kruthoff2017, Khalaf2018, Cano2018, Song2018, Wieder2020}. Higher dimensionality warrants the latter with stronger robustness against the material defects and perturbations. 
		
		Despite the advances in the field, the formation and stability of hinge modes in response to local defects or geometric features of the material remain subjects of ongoing debate. An investigation of these aspects will yield essential insights into the nature of topological states and advance the development of quantum devices based on these novel topological phases.
		Theoretical predictions have been made for switching on and off the hinge states, via $e-e$ interactions and Rashba spin-orbit coupling~\cite{Strom2010} or via the quantum confinement~\cite{Zhou2008}. But so far, the experimental evidence is still lacking. 
		On the other hand, the potential to host Majorana zero modes (MZMs) in helical hinge states via proximity-induced superconductivity has attracted considerable interest~\cite{Karzig2017, Fu2009, Hsu2018}. In a Josephson junction configuration, the MZMs give rise a $4\pi$-periodic contribution to the supercurrent as a function of the phase difference between the two superconducting leads~\cite{Fu2009}
		. This can be measured as doubled Shapiro steps as a response to the high-frequency signal~\cite{San-Jose2012,Houzet2013,Badiane2013,Bocquillon2016}. Recently, a switching current measurement in a bismuth nano-ring demonstrates the parity protection of the helical hinge states~\cite{Bernard2023}.

		While the topological nature of the hinge states is revealed by the $4\pi$-periodic supercurrent, its spatial distribution - localization on the edge - can be detected by the superconducting quantum interference (SQI) of the critical current $I_c(B)$. An enhanced supercurrent density $J_c$ at the edge of the flake results in an interference pattern similar to the superconducting quantum interference device (SQUID), different from the standard Fraunhofer-like pattern obtained for the homogeneous current distribution~\cite{tinkham2004introduction,Wang2021}. Such an unusual feature was reported previously in various systems that also identified with topological edge states, including the 2D TI~\cite{Bocquillon2016, Hart2014} and 3D SOTIs~\cite{Schindler2018Exp, Schindler2018Th, Li2014,Murani2017,Li2020,Chu2023,Choi2020, Kononov2020,Wang2020}.
		\begin{figure*}
			\centering
			\includegraphics[width=1\linewidth]{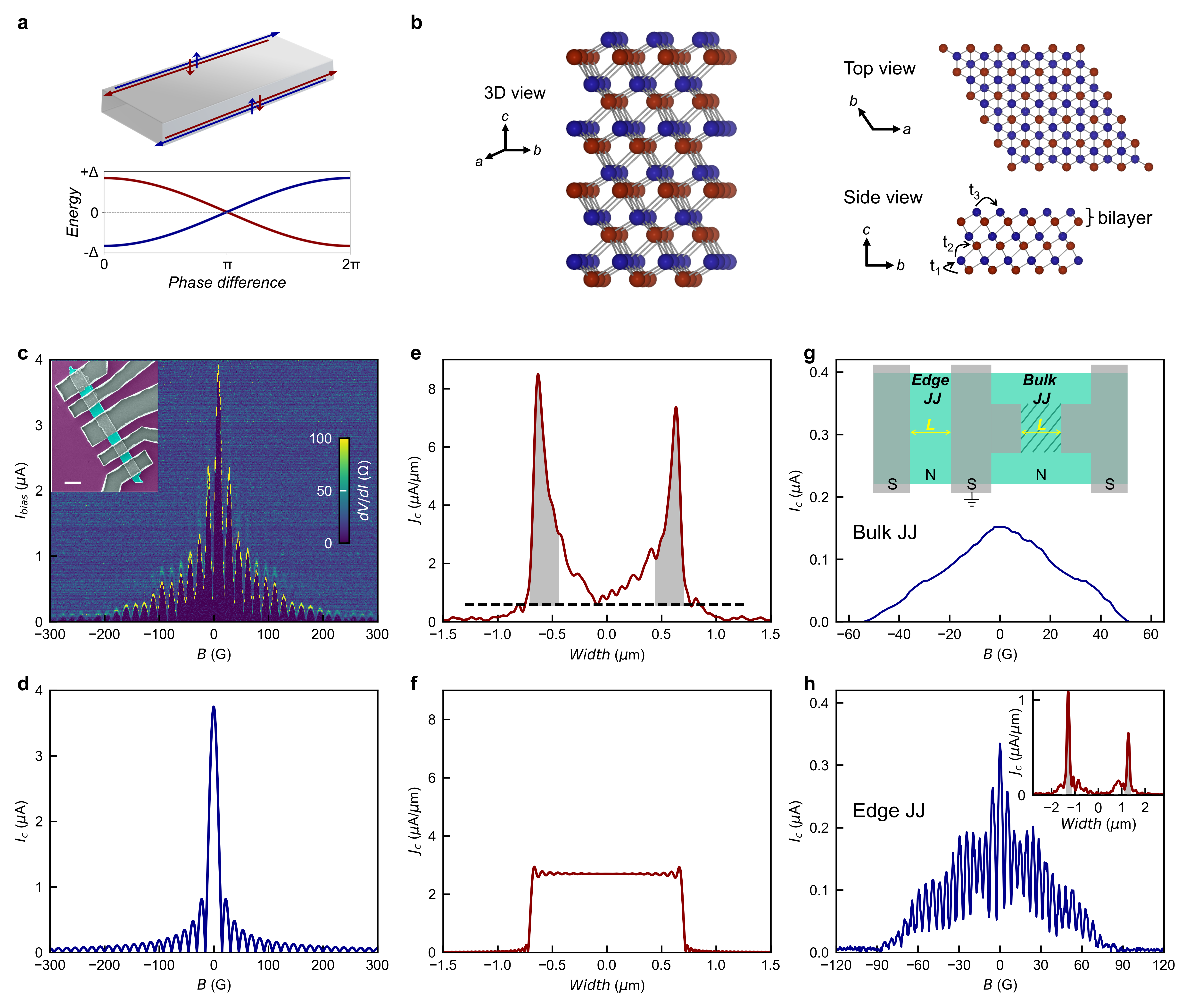}
			\caption{
				\textbf{Evidences of edge supercurrent in \BS Josephson junctions.} 
				\textbf{a}, \textit{left}: Schematic representation of multiple helical 1D hinge modes with spin up (blue) and spin down (red) localized along the step-edges and terraces in a 3D second-order topological insulator (SOTI). The hinge modes can host exotic $4\pi$-periodic Andreev bound states (Majorana bound states). In contrast, the bulk states host $2\pi$-periodic Andreev bound states, with a gap opening due to the lack of topological protection. \textit{right}: Classification of various material phases of Bi$_{1-x}$Sb$_{x}$ alloy based on Sb doping percentage. Experimental observations have confirmed the existence of higher-order hinge states in all three phases- Bi: semimetal~\cite{Schindler2018Exp}, Bi$_{0.97}$Sb$_{0.03}$: Dirac semimetal (this work), and Bi$_{0.92}$Sb$_{0.08}$: topological insulator~\cite{Aggarwal2021}.
				\textbf{b}, Schematic top-view drawing of sample F1: the bulk JJ (on the top surface, avoiding edges) and the edge JJ (spanning the full width of flake, including edges). The active junction area in bulk JJ is indicated by diagonal hatching. $S$: superconductor (Nb), $N$: normal region (Bi$_{0.97}$Sb$_{0.03}$). Both junctions have length $L=800$~nm.  
				\textbf{c}, Simulated Fraunhofer pattern (FP) for a JJ with homogeneous current distribution, following $I_c = I_{c,\mathrm{max}} \left| \mathrm{sinc}\left(\phi/\phi_0\right) \right|$, where $\phi = BL_{\mathrm{eff}}W$ is the magnetic flux through the junction and $\phi_0$ is the magnetic flux quantum. Inset shows the extracted $J_c(x)$ profile corresponding to the conventional FP, calculated from $I_cB$ using the Dynes–Fulton method~\cite{Dynes1971}, indicating the uniform current distribution with no edge enhancement.
				\textbf{d, e}, Comparison of $I_c(B)$ interference patterns for bulk and edge JJs on flake F1. The edge JJ (\textbf{d}) exhibits SQUID-like oscillations, while the bulk JJ (\textbf{e}) shows a monotonically decaying non-oscillating $I_c(B)$ pattern. Inset in \textbf{d} depicts the extracted $J_c(x)$ for the edge JJ, with strongly enhanced edge currents and negligible bulk contribution. Inset in \textbf{e} depicts a "gaussian-like" $J_c(x)$ peaked at the center, suggesting (quasi-) 1D ballistic transport in confined geometries.
				}

			\label{fig:fig1}
		\end{figure*}
		
		In this work we focus on Bi$_{1-x}$Sb$_{x}$ -- a large-scale compatible, air-stable, and free of volatile or toxic elements system. Depending on the doping, $x$, its phase ranges from semimetal to topological insulator. We show that when the system is tuned to the Dirac semimetal phase at 3\% Sb doping, it exhibits robust hinge states revealed by enhanced edge supercurrent. We compare the results to a tight-binding model, and find out that multiple hinges are formed due to the natural cleave edges of the Bi$_{1-x}$Sb$_{x}$ crystal. 
		Each hinge can host a SOTI channel.
		The high-frequency response not only shows a fractional Shapiro steps up to the second order, but also demonstrates its robustness in temperature and for the first time, a strong correlation with the existence of the hinge modes, together indicating the topological nature of the hinge modes. 
		Strikingly, we also find a series of 1D Rashba states, strongly localized at the edges. They turn out to be responsible for the broadening of the commonly observed edge current.

		\section{Results and discussion}
		\subsection{\textbf{Bulk-edge junctions: quasi-1D transport}}
		
		\noindent Bi$_{0.97}$Sb$_{0.03}$ flakes are exfoliated along the (111) plane. We fabricated twelve Josephson junctions (JJs) with lengths $L$ ranging from 300 to 1000~nm and BiSb flake thicknesses $t$ varying between 50 and 250~nm. Since the $4\pi$-periodic Josephson supercurrent in long-ballistic junctions can exceed the conventional $2\pi$-periodic supercurrent~\cite{Beenakker2013}, and diffusive contributions are suppressed in this regime, all JJs were designed to be in the (medium-to-)long junction regime. Here the junction length $L$ exceeds the superconducting coherence length $\xi_s$, which enhances the ratio of the $4\pi$-periodic contribution to the total Josephson supercurrent. In the main text, we focus on the results of the junctions on flakes F1 and F2. All the measurements were carried out at the base temperature 70 mK, unless indicated otherwise. Details for the other devices are provided in the Supplementary Material.
		
		To first demonstrate the strongly enhanced edge supercurrent, we fabricated two junctions on a same Bi$_{0.97}$Sb$_{0.03}$ flake F1: one in which the superconducting leads cross the entire junction width (edge JJ), and another where the leads are confined near the center of the junction, leaving the edges largely uncovered (bulk JJ). Both junctions have identical lengths of $L = 800$~nm. A schematic of the junction geometries is shown in Fig.~\ref{fig:fig1}b. The measured $I_c(B)$ patterns, shown in Fig.~\ref{fig:fig1}d and \ref{fig:fig1}e for the edge and bulk JJ, respectively, reveal two interesting features. First, the edge JJ exhibits a pronounced SQUID-like interference pattern indicative of strong edge supercurrent (Fig.\ref{fig:fig1}d). The real-space supercurrent density $J_c(x)$, calculated from the measured $I_c(B)$ using the Dynes and Fulton method~\cite{Dynes1971} (see Supplementary Section~S3 for details), shows a dominant edge contribution. Second, the bulk JJ shows a non-oscillating $I_c(B)$ response instead of a standard FP, indicating a quasi-1D transport (Fig.\ref{fig:fig1}e).
		
		The enhanced edge supercurrent can be related to the SOTI states forming at the hinges. However, from the $J_c(x)$ profile, we estimate the $I_c$ at two sides to be $I_{c,L} = 130$~nA and $I_{c,R} = 80$~nA, respectively, exceeding the theoretical maximum for a single helical mode for SOTIs~\cite{Choi2020, Chu2023}, suggesting that there are multiple edge channels contributing to the supercurrent.
		The atypical shape of the bulk JJ’s interference pattern may stem from the bulk quantization effects in the relatively small Bi$_{1-x}$Sb$_{x}$ flake ($W \sim 2300$~nm)~\cite{Fuseya2018, Ohtsubo2016}. Similar phenomenon of a few quasi-1D ballistic modes has been observed in quantum point contact devices of InAs-based heterostructure and graphene~\cite{Amado2013, Kraft2018}. We attribute this unusually large-scale quantum confinement effect to the extremely small L-pocket in Bi$_{0.97}$Sb$_{0.03}$. We will discuss these two points further in this report.

		\FloatBarrier
		\subsection{\textbf{Ballistic junction - $I_c(T)$}}
		\begin{figure*}
			\centering
			\includegraphics[width=0.9\linewidth]{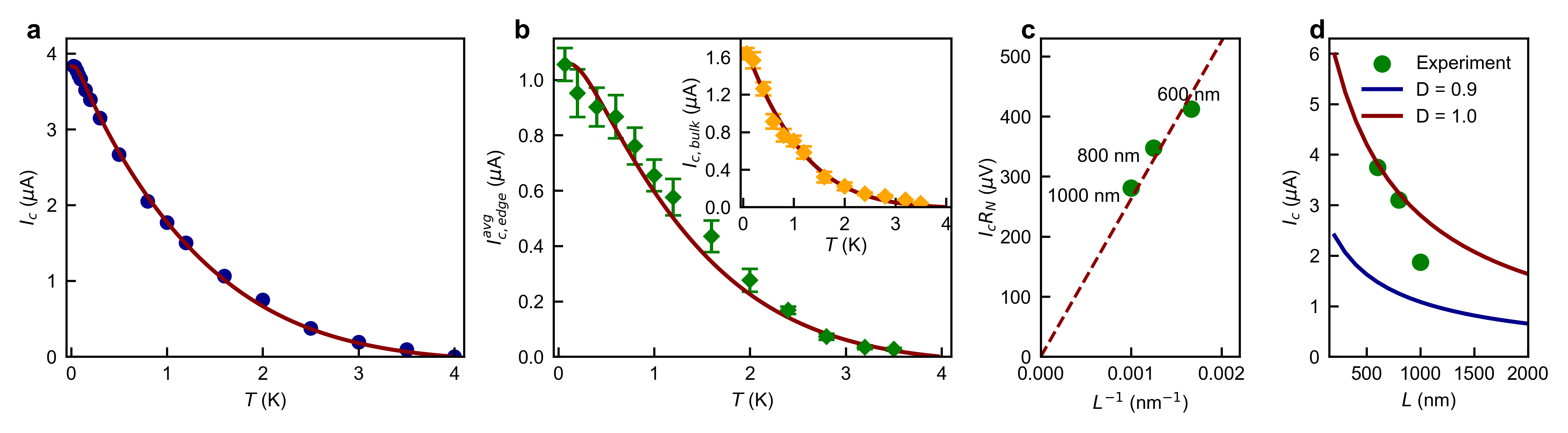}
			\caption{
				{\bf Ballistic transport in Bi$_{0.97}$Sb$_{0.03}$ junctions.} 
				\textbf{a}, Quantum interference pattern recorded at 70~mK for the F2$\_600$ junction, displaying SQUID-like oscillations in $I_c(B)$. The colormap represents the differential resistance $dV/dI$. Inset: SEM image of the flake F2 with four junctions of lengths 600, 1000, 900, and 800~nm (from top to bottom). Scale bar: 2~$\mu$m. Color scheme matches Fig.~\ref{fig:fig1}b.
				\textbf{b}, Extracted $J_c$ distribution across the junction width, calculated from \textbf{a} 
				, showing enhanced edge supercurrent density—characteristic of a SQUID-like interference. Contribution from hinge modes is shown by the gray shaded area. Residual bulk supercurrent is indicated by the black dashed line.
				{\bf c}, Temperature dependence of $I_c$ for device F2$\_600$. Measured data points (dark blue) can be fitted with Eilenberger model (solid dark red) for ballistic junctions with near-unity interface transparency. The estimated clean limit superconducting coherence length $\xi_s$ is 240 nm, which provides $L/\xi_s=2.5$. {\bf d}, Temperature dependence of the extracted $I_c$ for the edge modes (green diamonds) from $J_c$ distribution using the method described in Supplementary Section~S2. The inset shows the same for extracted bulk $I_c$ (orange diamonds). Both edge and bulk $I_c(T)$ can be fitted with Eilenberger model (solid dark red). The error bars are estimated from the standard deviation between experimentally and numerically calculated $J_c$ distribution as described in Supplementary Section~S1. {\bf e}, $I_cR_N$ scaling with inverse of junction length for different devices on flake F2 at 70~mK. The red dashed line highlights the linear trend. 
				{\bf f}, $I_c$ scaling with junction length for same devices as in \textbf{e}, where experimental values (green dots) lie within the Eilenberger framework with upper (solid dark red) and lower (solid dark blue) bounds set by the interface transparency D = 0.9 and 1.0, respectively.
			}
			\label{fig:fig2}
		\end{figure*}
		
		\noindent To further understand the transport properties of the junctions, we investigate the temperature dependence and the length dependence of the junctions. We present the results of the junctions with various lengths on another flake (F2). The inset of Fig.~\ref{fig:fig2}a shows a false-color scanning electron microscopy (SEM) image of the flake F2 (width $W = 1.4~\mu$m, thickness $t = 250$~nm), which hosts four JJs with lengths of 600, 800, 900, and 1000~nm. Figure~\ref{fig:fig2}a shows the magnetic field dependence of the critical current $I_c(B)$ for the 600~nm long junction. A pronounced SQUID-like modulation is observed, with all lobes exhibiting nearly equal widths of $\Delta B \simeq 18$~G. This corresponds to the magnetic flux quantum threading the effective junction area ($W \cdot L_{\text{eff}}$), where $L_{\text{eff}} \approx 900$~nm accounts for flux focusing effects due to the Nb electrodes. The calculated supercurrent density distribution $J_c(x)$, shown in Fig.~\ref{fig:fig2}b, reveals a pronounced enhancement of $J_c$ at both edges of the junction. Contrary to the intuitive expectation of a spatially uniform bulk supercurrent, the edge supercurrent in Fig.~\ref{fig:fig2}b smoothly decays into the bulk. Using our superconducting quantum interference (SQI) model, we attribute this spatial profile to different suppression rates for the bulk and edge current by the magnetic field. This leads to more robust oscillations at a higher field. Furthermore, a phenomenological smoothing effect—arising from finite coherence within the junction—modifies the extracted profile, meaning that the calculated $J_c(x)$ profile does not precisely reflect the zero-field current distribution (see details in Supplementary Section~S1). Unlike previous reports~\cite{Choi2020, Hart2014, Chu2023, Li2020}, the edge current profile in our data cannot be fitted by a simple Gaussian function. To consistently quantify the edge contribution, we develop a systematic procedure to determine, what we dub as the ‘optimal width’— the spatial extent of the edge states that yields the best agreement between the extracted and actual supercurrent density. Details of the extraction and optimization procedure are provided in Supplementary Section~S1. In Fig.~\ref{fig:fig2}b, the shaded gray regions indicate the edge supercurrent contributions for the optimal width, while the dashed line denotes the residual bulk supercurrent. For the F2$\_600$ junction, the extracted edge width is approximately 250~nm on each side, yielding left and right edge supercurrents of $I_{c,L} = 1.19~\mu$A and $I_{c,R} = 0.93~\mu$A, respectively, and a bulk supercurrent of $I_{c,\text{bulk}} = 1.64~\mu$A.
		
		The temperature dependence of the critical current $I_c(T)$ is shown in Fig.~\ref{fig:fig2}c (dots). The $I_c$ exhibits an almost non-saturating increase at low temperatures, which is indicative of a highly transparent interface~\cite{Heida1998}. Previous studies on Bi$_{0.97}$Sb$_{0.03}$ have shown that this material hosts high-mobility carriers with exceptionally long mean-free paths~\cite{Li2018}, suggesting that our junctions are in the ballistic regime.
		To analyze the data quantitatively, we fit the measured $I_c(T)$ using the Eilenberger formalism developed by Galaktionov and Zaikin~\cite{Galaktionov2002}. The fit (solid line in Fig.~\ref{fig:fig2}c) corresponds to a junction with $L = 600$~nm and yields the following parameters: interface transparency $D = 0.9995$ (assumed equal for both left and right interfaces), and normal-state resistance $R_N = 208~\Omega$, corresponding to 62 modes. The superconducting coherence length is estimated to be $\xi_s = 240$~nm, taking $T_c \sim 4$~K from the measurement and $v_F\sim 8\times 10^5 $~m/s as an effective Fermi velocity.  
		It is important to note that the $T_c$ of the junction is lower than the intrinsic $T_c$ of Nb ($\sim 9$~K), consistent with the proximity-induced superconducting gap in the Bi$_{1-x}$Sb$_{x}$ flake beneath the Nb contacts. In this configuration, proximitized Bi$_{1-x}$Sb$_{x}$ regions act as superconducting leads, bridging the central normal Bi$_{0.97}$Sb$_{0.03}$ channel with fully transparent contacts. Similar high-transparency proximity effects in Bi$_{1-x}$Sb$_{x}$ have been reported in earlier studies~\cite{Li2018,Li2019}.
		The fitted $R_N$ value is notably higher than the experimentally measured $R_N = 22~\Omega$. This is due to the presence of multiple transport channels: while all channels contribute to the normal-state resistance, only a subset of them may carry supercurrent. The overall fit supports the conclusion that the junction is in the intermediately-long-ballistic regime ($\xi_s < L < l_e, L/\xi_s\sim 2.5$). Similar behavior is also observed in other devices (see Supplementary Sections~S4 and S5 for details).
		
		Based on the spatial current density $J_c(x)$, we extract the edge and bulk contributions to the supercurrent and analyze their temperature dependencies, as shown in Fig.~\ref{fig:fig2}d. For simplicity, an average edge supercurrent is defined as $\widetilde{I}_{c,\mathrm{edge}} = (I_{c,L} + I_{c,R}) / 2$. Both edge and bulk states $I_c(T)$ can be fitted separately using the same Eilenberger model. 
		For device F2$\_600$, the fitting parameters are: $D = 0.99$ ($0.9997$) 
        and $R_N = 637~\Omega$ ($535~\Omega$), corresponding to 20 (24) modes for the edge (bulk) components.
		
		In a ballistic junction, the characteristic energy is determined by the smaller one between the superconducting gap $\Delta$ and the Thouless energy $E_{Th}$. The time for charge carriers to flow through the junction is estimated to be $\sim L/v_F$, where $v_F$ is the Fermi velocity and $L$ is the junction length. Thereby, the Thouless energy $E_{Th}\sim \hbar v_F/L$~\cite{Ishii1970, BenShalom2015}. For a short-ballistic junction, the maximal critical supercurrent for one mode is given by $I_c = e\Delta/\hbar$~\cite{Beenakker2013}, which equals $148$ nA for $T_c = 4$~K. From the Eilenberger fitting, we estimate a critical current $I^0_c$ of $52$~nA for a single channel in the 600~nm junction. Thus, the average number of channels per edge can be estimated to be $1\mu A/52~ $nA$\sim $ 20 modes. A similar estimation can be done for the 800 and 1000 nm junctions on flake F2, where the number of channels per edge is 6 and 8 modes, respectively (see Supplementary sections~S4 and S5). Note that the number of modes is estimated by considering all channels to be normal spin-degenerate 2$\pi$ modes. If these are 4$\pi$ modes in an intermediate long junction, the number of modes should be reduced by a factor between 1 and 2. When a junction is in the long-junction regime, the product $I_cR_N\propto1/L$. In Fig.~\ref{fig:fig2}e, we plot the $I_cR_N$ products of all three junctions on F2, which match well to the linear trend. To demonstrate the generally high transparency of the junctions, we plot the critical currents $I_c$ of all three junctions on the same BiSb flake as a function of junction length $L$ in Fig.~\ref{fig:fig2}f, alongside simulated curves based on the Eilenberger theory. All data points lie between the theoretical curves corresponding to interface transparencies $D = 0.9$ and $D = 1$, indicating that the $I_c$ is dominated by the consistently high transparency modes across all devices.
		
		\subsection{\textbf{Topological protection - 4$\pi$ supercurrent correlation}}
		
		\noindent To verify the nontrivial topological nature of the hinge modes, we performed radio-frequency (RF) measurements to investigate the Shapiro steps. In conventional Josephson junctions with a $2\pi$-periodic current–phase relation (CPR), the voltage steps are quantized as $V = nhf/2e$, where all integer steps appear. In contrast, in the presence of topologically protected gapless Majorana zero modes (MZMs), the CPR becomes $4\pi$-periodic, resulting in the suppression of odd Shapiro steps, a hallmark of the fractional a.c.\ Josephson effect~\cite{Fu2009}. The observation of missing odd steps, particularly beyond the first one, has been reported as a characteristic signature of a topological $4\pi$-periodic supercurrent~\cite{Li2018, Rokhinson2012, Wiedenmann2016, Bocquillon2016}. Nevertheless, it is also recognized that a $4\pi$-periodic CPR may arise in non-topological scenarios. For example, specific parameter regimes of the resistively, capacitively and inductively shunted junction (RCLSJ) model can mimic such behavior~\cite{Liu2025}, and extremely high junction transparency can also lead to similar effects in trivial systems~\cite{Shabani2021}. Therefore, it is crucial to demonstrate a direct correlation between the observed $4\pi$-periodic signals and the underlying topological states in a wide-range parameter space.
		
		\begin{figure*}
			\centering
			\includegraphics[width=1\linewidth]{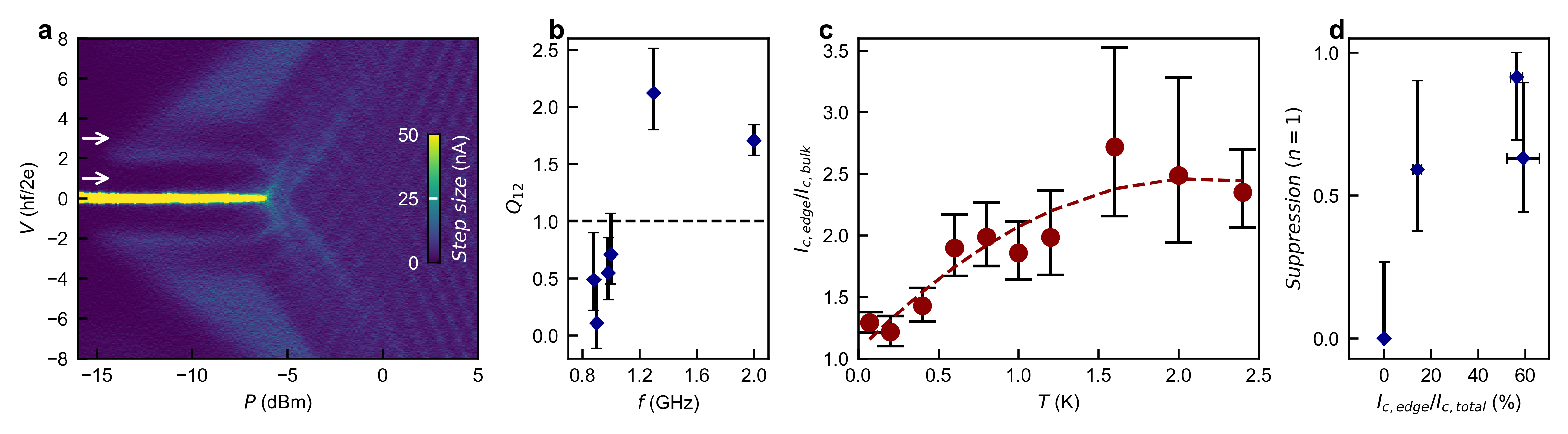}
			\caption{
				{\bf Fractional a.c. Josephson effect and topological protection of the 1D hinge modes.} 
				{\bf a}, Shapiro steps in device F2$\_600$ measured under radio-frequency (RF) excitation of 0.9~GHz at 70~mK. The colormap shows the Shapiro step size calculated from the d.c. voltage bins as a function of d.c. voltage normalized to $\frac{hf}{2e}$ and RF power. White arrows mark the missing odd ($n = 1, 3$) Shapiro steps. {\bf b}, Frequency dependence of the ratio between $n=1$ and $n=2$ step size ($Q_{12}$). 
				{\bf c}, Suppression factor ($1-Q_{12}/Q_{12}^{max}$) as a function of the edge-mediated supercurrent for different devices, highlighting an important observation that the $n=1$ Shapiro step is more diminished in junctions with higher edge state contribution. This suggests overall topological protection of the 1D hinge modes.
				{\bf d}, $I_{c,\ edge}/I_{c,\ bulk}$ ratio for device F2$\_600$ showing an increasing trend with temperature up to 2K, suggesting that the bulk $I_c$ decays more rapidly than that of the ballistic edge states. The dashed curve is guide-to-the-eye. 
			}
			\label{fig:fig3}
		\end{figure*}

		In Figure~\ref{fig:fig3}a, we show the Shapiro step binning map for device F2$\_600$ under 0.9~GHz irradiation, where the first and third steps ($n = 1$ and $3$) are absent, as indicated by white arrows.
		The missing-step phenomenon is measurable from 0.6 GHz, and these steps reappear at higher microwave frequencies~(Supplementary Section~S6). The suppression of the odd steps can be phenomenologically quantified by the $Q_{i,i+1}$ factor, 
		where $i$ is an odd number. $Q_{i,i+1}$ represents, in general, the ratio between the width of an odd step and that of its successive even step. When $Q_{i,i+1}<1$, step $i$ is considered as reduced. In Fig.~\ref{fig:fig3}b, factor $Q_{12}=w1/w2$ is plotted as a function of the frequency (see Supplementary Section~S13 for estimation details). The observed frequency dependence rules out alternative explanations such as Landau-Zener transitions~\cite{Wiedenmann2016}. Moreover, at 0.9~GHz, the odd-step suppression remains visible up to 1.2~K (Supplementary Section~S7), underscoring the thermal stability of the $4\pi$-periodic component. 
		
		Fractional Shapiro steps are observed in multiple devices (Supplementary Sections~S8 and~S9), spanning a broad range of RCSJ model parameters (\( R \in [20, 30]~\Omega \), \( I_c \in [0.3, 3.8]~\mu\text{A} \)). This indicates that the missing steps cannot be simply attributed to a specific choice of RCSJ parameters.
		
		As an important step, we test the correlation between the presence of the edge state and the visibility of the fractional Shapiro steps. We define a normalized suppression factor $\gamma = 1-Q_{12}/Q_{12}^{max}$ for all junctions. $Q_{12}^{max}$ is the maximal value of $Q_{12}$ among all measured junctions and frequencies. If $\gamma =1$, it means that the first step is fully suppressed; oppositely, if $\gamma=0$, there is no suppression of the odd steps.
		In Fig.~\ref{fig:fig3}c, we plot the suppression factor $\gamma$ as a function of the estimated percentage of the edge supercurrent in various junctions. A positively correlated behavior can be found clearly, suggesting that the hinge states are the topological origin of the observed fractional Shapiro steps.
		
		As we demonstrated before, both the bulk and edge channels are ballistic. However, the hinge modes, due to their topological protection, are expected to exhibit enhanced robustness against thermal noise. We assess this robustness by examining the temperature dependence of the ratio between edge and bulk critical currents, defined as $\alpha = I_{c,\ \mathrm{edge}} / I_{c,\ \mathrm{bulk}}$, as shown in Fig.~\ref{fig:fig3}d, where $I_{c,\mathrm{edge}} = (I_{c,L} + I_{c,R})$. For device F2$\_600$, $\alpha$ displays an increasing trend up to 2~K, suggesting that the edge modes maintain their coherent transport while thermal effects increasingly suppress the bulk contribution. Such behavior has been previously linked to the topological protection of edge states, whereby the bulk transitions toward a diffusive regime with increasing temperature, while the topological hinge channels preserve their ballistic nature~\cite{Schueffelgen2019}. Similar trends in $\alpha$ were consistently observed across other devices (Supplementary Sections~S4 and S5), providing compelling evidence for the existence of topologically protected 1D hinge modes in Bi$_{0.97}$Sb$_{0.03}$.

\begin{figure*}
	\centering
	\includegraphics[width=0.8\linewidth]{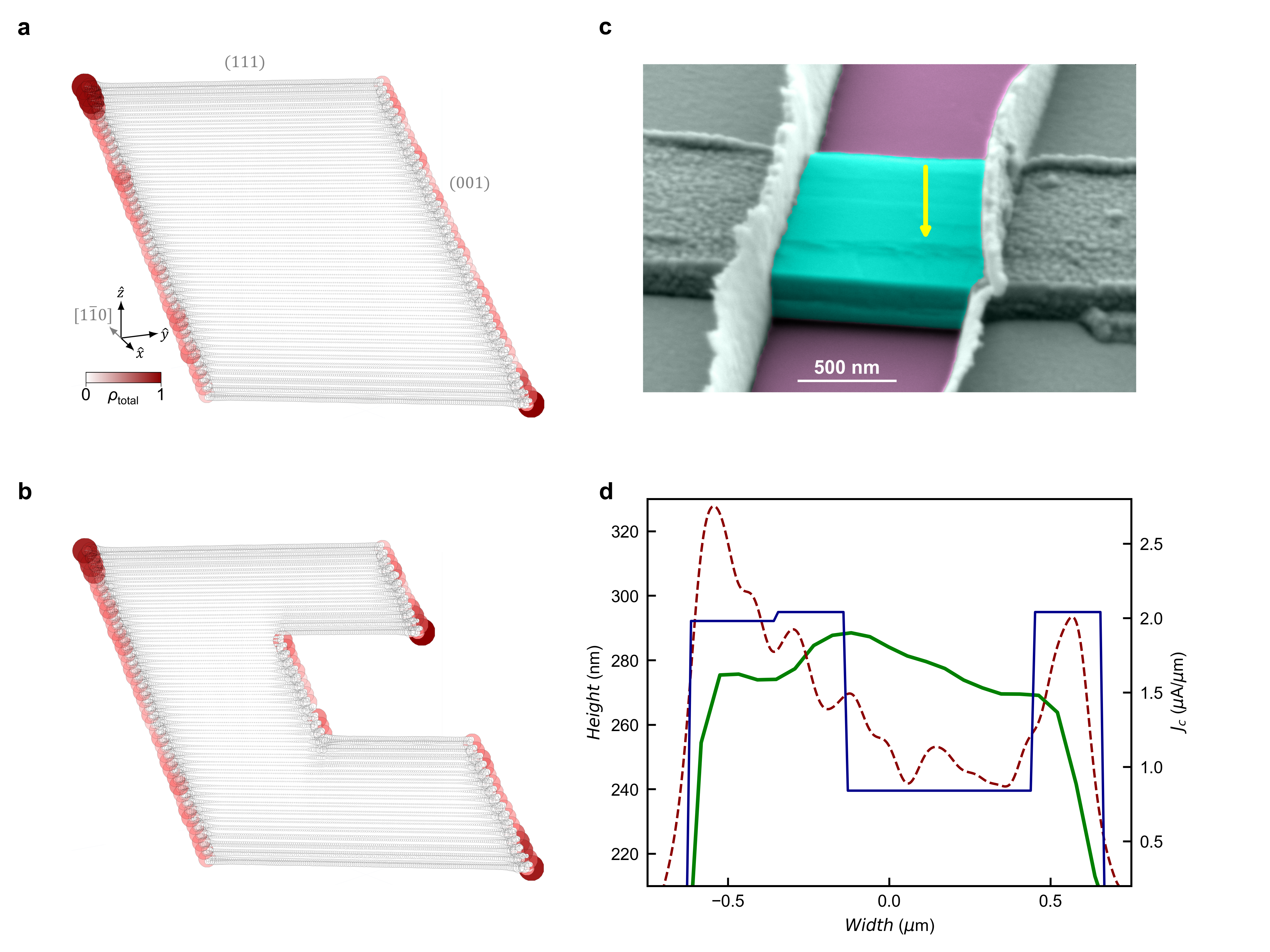}
	\caption{
		{\bf Experimental and theoretical signatures of the effect of multiple structural irregularities on hinges.} 
		{\bf a}, Cross-sectional SEM image of the F2$\_1000$ device showing the additional step (black arrow) present near one of the junction edge (also visible in AFM image, Supplementary Section~S11). 
		{\bf b}, Overlayed AFM height profile (solid turquoise) showing the $\sim15$~nm high step near the left edge of the junction, designed $J_c^{\text{set}}(x)$ model (shaded region) including this step (black arrow) and experimental $J_c^{\text{exp}}(x)$ distribution (dashed darkred) showing the enhanced width of left edge due to this step. $J_c^{\text{set}}(x)$ is scaled as per the $J_c^{\text{exp}}(x)$ data.
        {\bf c}, Tight-binding band structure of a 20 bilayer Bi$_{0.97}$Sb$_{0.03}$ slab (8 nm thick by 16 nm wide) with translation symmetry in the $[1\overline10]$ direction, confinement along the top $(111)$ and side $(11\!-\!2)$ surfaces, and with four identical artificial step edges on one side. The color intensity represents the localization defined as the sum of the wavefunction density over the 10 sites around the site where the wavefunction density is maximal. The most strongly localized bands are four times degenerate (Rashba).
		{\bf d}, The summed wavefunction density of the highlighted degenerate Rashba (blue) and SOTI (red) modes at the Fermi level $(E_\text{F}=0)$.
	}
	\label{fig:fig4}
\end{figure*}

\subsection{\textbf{Multiple hinge modes pinned to structural irregularities}}         
\noindent We now turn our attention to the multi-edge channels. As discussed previously, the observed supercurrent must be carried by multiple (broadened) edge channels, which appears to deviate from the SOTI prediction of a single hinge mode per edge. A plausible explanation lies in the structural imperfections of exfoliated flakes: side edges are rarely atomically flat or perfectly aligned with a single crystallographic facet. Instead, steps, discontinuities, and other irregularities often form along the edge, effectively introducing additional internal boundaries that can host localized hinge modes~\cite{Nayak2019, Sekine2023}.

We directly observe this effect in a junction with an additional step. In Fig.\ref{fig:fig4}a, a scanning electron microscope (SEM) image shows the side view junction F2\_1000. A 15 nm step is found on the top surface. To reproduce the measured $I_c(B)$ of this junction, it is necessary to include both the step feature and an additional hinge current density at the step in the model, resulting a strong asymmetric current distribution (Fig.\ref{fig:fig4}b, shaded area). This is qualitatively in good agreement with the extracted supercurrent profile (Fig.\ref{fig:fig4}b, red dotted line).

\begin{figure*}[!tbh]
			\centering
			\includegraphics[width=1\textwidth]{Figures/Figure_5.png}
			\caption{
				{\bf Finite-size effect on the SQI pattern.}
				{\bf a-c}, Comparing the $I_cB$ for {\bf a}, $W = 1800$ nm (300 mK), {\bf b}, $W = 700$ nm (100 mK) and {\bf c}, $W = 300$ nm (100 mK) wide junctions reveal the quantum confinement effect due to width of an edge JJ. A gradual shift from the SQUID-like response in widest flake to a monotonically decaying $I_cB$ in the narrowest flake was observed.
			}
			\label{fig:fig5}
		\end{figure*}

To investigate this phenomenon, we perform tight-binding (TB) calculation using Kwant~\cite{Groth2014} on the Bi$_{0.97}$Sb$_{0.03}$ crystal, based on the 16-band tight-binding model of Liu and Allen~\cite{Liu1995}. We construct a slab geometry with translational symmetry along the $[1\bar{1}0]$ direction and confinement along the top $(111)$ and side $(11\! - \!2)$ surfaces. Structural irregularities are introduced by forming four identical step edges along the $(11\! - \!2)$ surfaces.
The resulting band structure is shown in Fig.\ref{fig:fig4}c, where the highlighted bands are strongly localized at the hinges and their wave function density at the Fermi level are plotted in Fig.\ref{fig:fig4}d. This confirms our idea that the steps can activate new hinge modes.

However, this alone cannot fully account for the generally broadened edge current profile (approximately 100--200~nm). From the simulations, one can see that the activated HOTI states remain well localized within only a few nanometers from the hinge. A closer inspection of the band structure in Fig.\ref{fig:fig4}c reveals that
the set of localized hinge states contains (i) a pair of opposite spin Rashba modes, both connecting only conduction band and (ii) a single spin SOTI mode, connecting conduction and valance bands. 
Both states are 4-fold degenerate due to the identical step structure, which is chosen to keep the dispersion clean and to directly relate the observed degeneracy to the number of steps.

The coexistence of both Rashba and SOTI 1D states is only visible when a real crystal lattice is taken into account. This leads to a clearer interpretation of the commonly observed broadened edge current: Both states can carry supercurrent. A Rashba mode at momentum $k$ can back-scatter to the state at $-k+\delta k$ which has the same spin and is localized on the same hinge, directly leading to the delocalization of the states and therefore a broadening of the edge current. Such Rashba-type states localized at the edge have been theoretically proposed~\cite{Smoluchowski1941,Xu2018} and experimentally observed in STM and ARPES measurements on Bi-based systems~\cite{Fedotov2017,Ko2024,Takayama2015}. However, distinguishing between the helical and Rashba states has remained challenging in these types of experiment.
Although both types of states can contribute to the edge supercurrent, the SOTI states are topologically protected, whereas the Rashba states are not. Consequently, only the SOTI states are relevant to the observed $4\pi$ component. From the frequency dependence, the $4\pi$ current fraction is estimated to be about 20\% , consistent with this picture, corresponding to only 2-4 SOTI channels among approximately 10-20 Rashba-type channels.

		Another peculiar feature that we have observed is the quasi-1D transport in the bulk states. To obtain a better understanding of the size effect, we carried out the measurement on junctions with different sizes. A striking width dependence was found. In Fig.~\ref{fig:fig5}a-c, we plot the $I_c(B)$ of three different junctions which across the entire flake (namely the edge JJ as defined previously) with widths W = 1800 nm, 700 nm and 300 nm, respectively. As the width decreases, the measured SQI pattern changes from a clear SQUID-like shape to a monotonic decay. This confirms our idea that in these long junctions, besides the hinge modes, the bulk supercurrent is carried by the bands with extremely small wave-vector $k_F$, most likely from the L-pocket, which are quantized even in the sub-micron size flakes.

		More importantly, the edge current disappears in the 300~nm narrow junction. This observation is consistent with the estimated edge width, suggesting that the extended edge states on the two sides can couple to each other, leading to the opening of an effective gap in the Rashba states. Although previous theoretical studies have suggested that the finite-size inter-coupling between opposite topological hinge modes can open a gap in the one-dimensional helical states~\cite{Zhou2008}, we argue that the calculated wavefunctions of all hinge states are strongly localized within only a few nanometers. Therefore, such coupling based on the direct overlap of hinge-state wavefunctions is unlikely. The apparent disappearance of the hinge-related supercurrent is instead mainly a consequence of the reduced transparency to the topological hinge states when most surrounding Rashba states become gapped.
		
		\section{Conclusion}
		\noindent 
		We have observed robust hinge states in Bi$_{0.97}$Sb$_{0.03}$ single crystals, with results that are highly reproducible. Importantly, our work demonstrates, for the first time, a direct correlation between the $4\pi$-periodic Josephson effect and the presence of hinge modes, thereby confirming their topological nature. Furthermore, both theoretical and experimental results show that multiple hinge states can be activated by the step-like structure of the crystal. This firmly establishes the Bi$_{1-x}$Sb$_x$ alloy system as a promising platform for advancing topological quantum devices, opening new opportunities for quantum network-based architectures through nano-engineering of artificial one-dimensional edges. The emergent one-dimensional Rashba states are identified as the primary origin of the extended edge current distribution. This not only provides a comprehensive explanation for the widely observed broadened edge profiles across various SOTI systems, but also opens new possibilities for designing and implementing topological devices that leverage Rashba-state–dominated edge transport. 
		
		\section*{Data availability}
		The experimental datasets that can be used to reproduce the findings of this study are available via Zenodo.
		
		\section*{Methods}
		\subsection*{\textbf{Crystal growth}}
		Bi$_{1-x}$Sb$_x$ single crystals are grown using a modified Bridgman method. High-purity Bi ingots (99.999\%) and Sb ingots (99.9999\%) were packed in a cone-shaped quartz tube and sealed under vacuum ($4\times10^{-7}$ mbar). The tube was first put in a box furnace and heated up to 600 $^{\circ}$C for 12 hours. The tube was shaken several times to obtain a homogeneous mixture of Bi and Sb. Then the tube was quickly cooled to room temperature and hung vertically in a mirror furnace for crystal growth. The tube was heated to 300-400 $^{\circ}$C, starting from the cone-shaped bottom, and the molten zone was translated up at a rate of 1 mm/hour. Flat crystals up to 1 cm in length were obtained by cleaving the crystal boule. 
		
		\subsection*{\textbf{Device fabrication}}
		We performed micro-mechanical exfoliation of Bi$_{0.97}$Sb$_{0.03}$ single crystals to transfer high-quality flakes on pre-cleaned Si/SiO$_2$ substrates with 120 nm oxide thickness. The thickness of the flakes was measured using a commercial atomic force microscope (Icon, Bruker). Josephson junctions were fabricated on the desired flakes using standard electron-beam lithography, followed by Ar$^+$ etching at 300~W RF power for 30~s to remove native oxide and contamination layers before in-situ sputter deposition of 120~nm Nb electrodes and 2~nm of Pd as a capping layer to protect the Nb from oxidation. Nb d.c. Sputtering was performed at a relatively slow rate, 10~nm/min, to mitigate the Ar plasma-induced degradation of the BiSb flakes.       
		
	\section*{Acknowledgments}
        \subsection*{Funding:}
        C.L. thanks the Netherlands Organization for Scientific Research (NWO) for the financial support through a VIDI grant (VI.Vidi.203.047) and the Gravitation program QuMat (024.005.006). 

        \subsection*{Author contributions:}
        B.B.: Methodology, software, validation, formal analysis, investigation, data curation, writing, visualization. S.d.W.: Software, validation, investigation, writing. Z.W.: Methodology, investigation, data curation, visualization. Y.K.H.:Methodology, validation. M.G.: Methodology, validation, investigation. A.B.: Conceptualization, methodology, validation, investigation, writing, and supervision. C.L.: Conceptualization, methodology, software, validation, formal analysis, investigation, writing, project administration, funding acquisition, and supervision.
        \subsection*{Competing interests:}
        The authors declare that they have no competing interests.
        
        \subsection*{Data and materials availability:}
        All data needed to evaluate the conclusions in the paper are present in the paper and/or the Supplementary Materials.
		
		\bibliographystyle{science}
		\vspace{1ex}
	
\bibliography{reference}
\end{document}


\maketitle

\begin{center}
\Large \textbf{Supplementary Information}
\end{center}

\tableofcontents  
\addtocontents{toc}{\setcounter{tocdepth}{1}} 

\begin{spacing}{2}

\newpage
\section{Superconducting quantum interference model}
\label{sect:sec1}  
\subsection{Simulation of the Field Dependence in SNS Junctions}

The field dependence of the SNS junction is simulated using a phenomenological model based on Ref.~\cite{Chen2018}. In this model, we introduce an effective decay length, \( L_c \). Instead of a \( 1/L \) dependence, we consider an exponential decay $J_c = J_0\exp(-L/L_c)$. When applying the Landau gauge, a phase is acquired as the carrier moves from \( x_1 \) to \( x_2 \), given by
$\frac{2\pi B_z (y_2 + y_1)}{\phi_0}$.

\begin{figure}[!tbh]
\begin{center}
\includegraphics[width=0.5\textwidth]{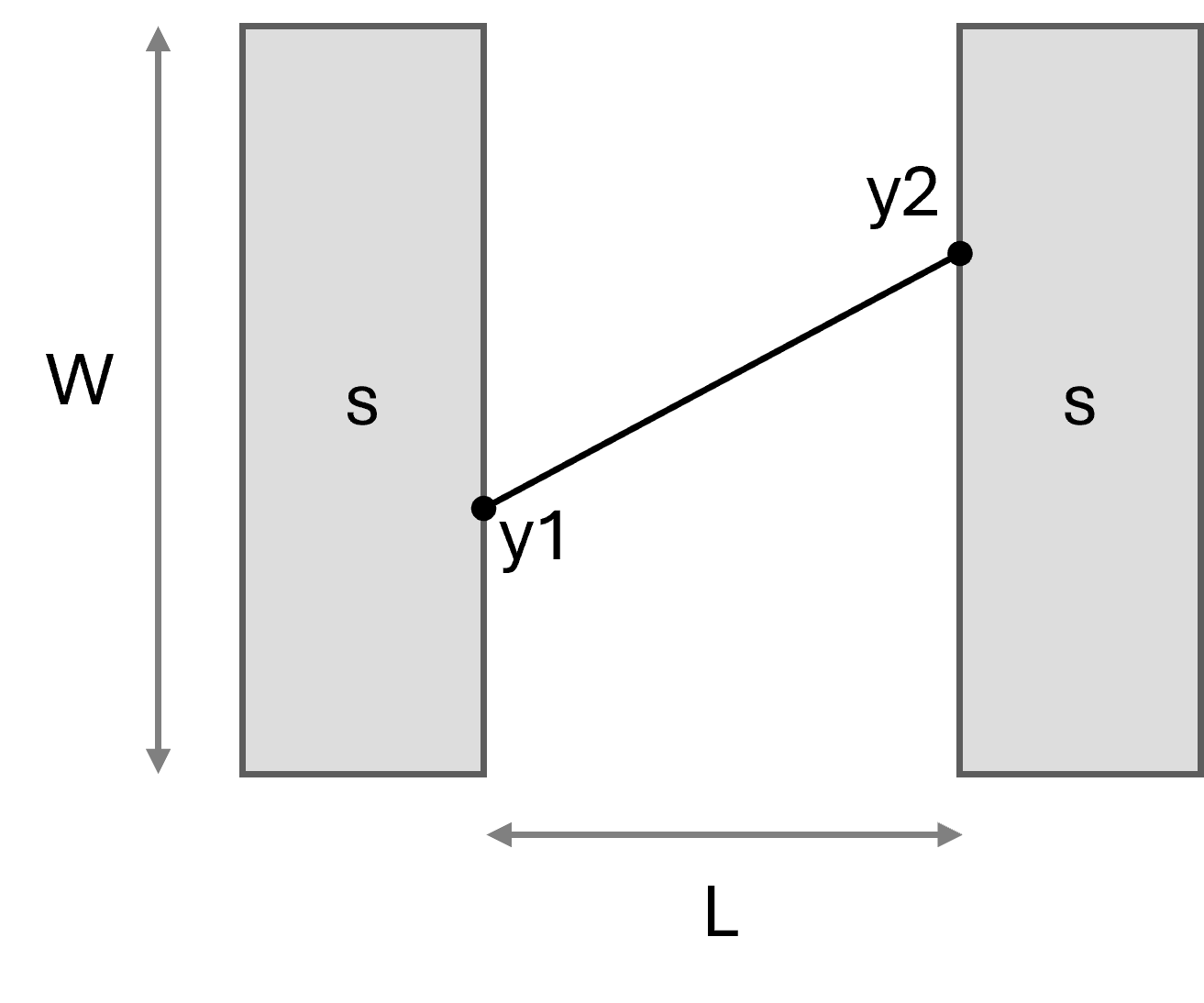}  
\end{center}
\caption{{\bf $|$} Schematic of the superconducting quantum interference model.} 
\label{fig:figS1-1}
\end{figure} 

The complete expression for the supercurrent is:

\begin{equation}
I(\Delta \phi_0, B_z) = \int_{0}^{W} \int_{0}^{W} \frac{1}{L^2 + (y_2 - y_1)^2} \sin(\Delta \phi(B_z)) \, dy_1 dy_2,
\label{eq:1.1}
\end{equation}

where the total phase gained by a particle traveling from \( (0, y_1) \) to \( (L, y_2) \) is given by:

\begin{equation}
\Delta\phi(B_z) = \Delta\phi_0 + \frac{2\pi B_z L (y_1 + y_2)}{\phi_0}.
\end{equation}

Here, \( L \) is the effective junction length, accounting for the magnetic field focusing effect, and \( \Delta\phi \) is the superconducting phase difference between the two leads. The Josephson current is obtained by varying \( \Delta\phi \) and maximizing Eq.~(\ref{eq:1.1}).

While refining the model, we identify three key findings:
\begin{enumerate}
    \item The resolution of \( J_c(x) \) depends on the decay length \( L_c \).
    \item The shape of the edge current can be smoothed, and in some cases, its width deviates from the actual distribution.
    \item The model captures various details, including steps and edge currents.
\end{enumerate}

\subsection{Procedure for Obtaining the Optimal Edge Width}

In this study, we resolve the edge states by comparing both \( I_c(B) \) and the converted \( J_c(x) \), obtained using the Dynes and Fulton method (explained in the next section), with simulation results. The analysis follows these steps:
1) Optimize the \( I_c(B) \) simulation by minimizing the standard deviation (std) between the simulated and experimental data.
2) Convert both the simulated and experimental data into \( J_c(x) \), then compute the std and uncertainties.
3) Extract the edge current percentage from the converted \( J_c(x) \) by varying the edge width and compare it to the original edge current percentage defined in the model.

\subsubsection{Optimization of \( I_c(B) \) for Simulation}

Based on the measured \( I_c(B) \), we compute \( J_c^{\text{exp}}(x) \). A simple \( J_c(x) \) model, consisting of both edge and bulk currents, is then designed. By tuning parameters such as edge width and bulk current density, we optimize the results by minimizing the std.

\begin{figure}[!tbh]
\begin{center}
\includegraphics[width=\textwidth]{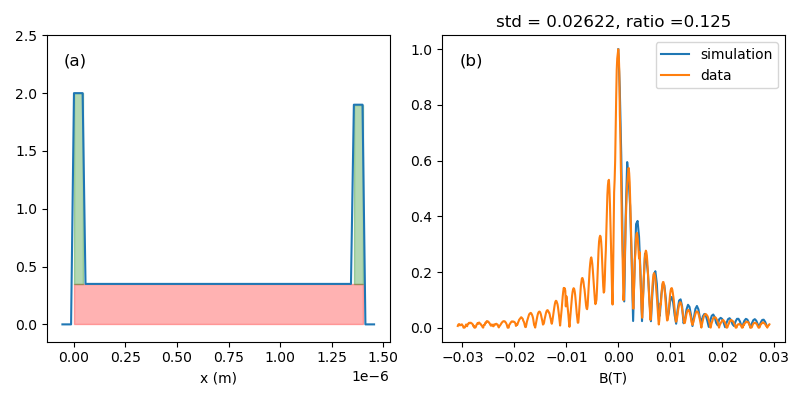}  
\end{center}
\caption{{\bf $|$} Optimization of the simulation for a 600 nm junction. (a) Designed current distribution: red represents bulk current, and green represents edge current. (b) Comparison between simulated and experimental \( I_c(B) \) data. The standard deviation is displayed on top.} 
\label{fig:figS1-2}
\end{figure} 

\subsubsection{Computation of \( J_c(x) \) and Uncertainty Estimation}

Once \( I_c(B) \) is obtained, the current distribution \( J_c(x) \) is calculated for both the simulation and experimental data. The std is then computed, and the uncertainty in current is determined by:
$\sigma = \text{std} \times I_{\text{tot}}$.

\subsubsection{Estimation of the Edge Width from the Simulation}

\begin{figure}[!tbh]
\begin{center}
\includegraphics[width=\textwidth]{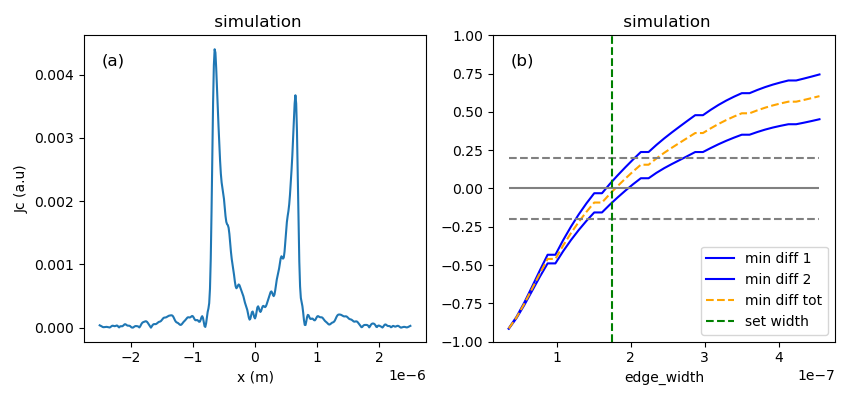}  
\end{center}
\caption 
{{\bf $|$} Estimation of the edge width.} 
\label{fig:figS1-3}
\end{figure}

A notable feature of the edge current is that the calculated current distribution appears smoothed. For instance, the computed \( J_c^{\text{sim}}(x) \) for the sharp edge current distribution shown in Fig.~\ref{fig:figS1-2}(a) is presented in Fig.~\ref{fig:figS1-3}(a).

The actual edge width is no longer easily discernible by eye. To address this, we compare the percentage of the edge current in the calculated \( J_c(x) \) at different widths (\( P_{\text{edge,sim}} \)) with the designed edge current percentage (\( P_{\text{edge,set}} \)) (Fig.\ref{fig:figS1-2}(a)). The procedure for determining \( J_c(x) \) is outlined in Section \ref{sect:sec2}.  
Ultimately, we plot the relative difference \((P_{\text{edge,sim}} - P_{\text{edge,set}}) / P_{\text{edge,set}}\), as shown in Fig. \ref{fig:figS1-2}(b). From this analysis, we identify an optimal edge width for a given set of parameters at approximately 250 nm, where the dashed orange line crosses zero. Notably, this optimal width is larger than the originally designed edge width (green dashed line). This estimation method is applied to all presented devices.

\subsection{Averaging Effect in Resolving Current Distribution}

During the simulation, we observed several effects related to the model that may be useful for broader applications, including its limitations.

\begin{figure}[!tbh]
\begin{center}
\includegraphics[width=\textwidth]{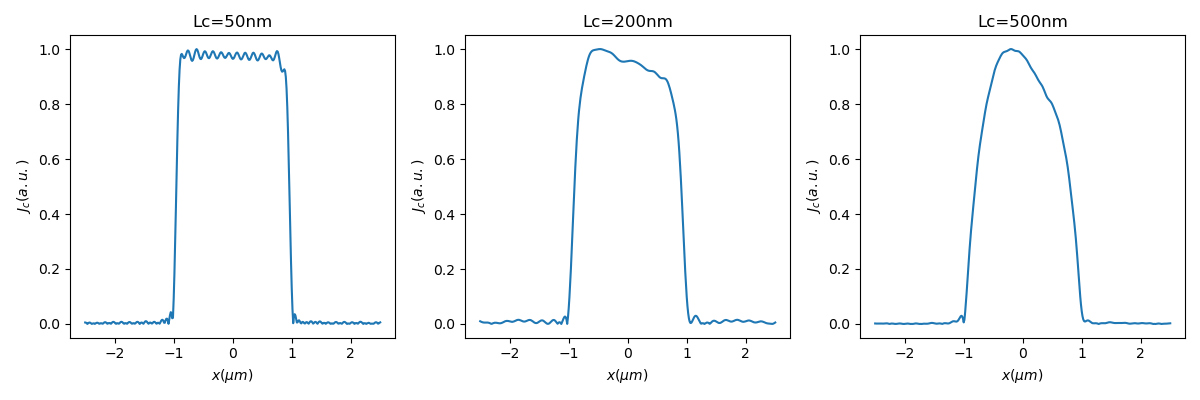}  
\end{center}
\caption{{\bf $|$} Estimation of the edge width.} 
\label{fig:figS1-4}
\end{figure} 

The effect of the decay length \( L_c \) is illustrated in Fig.~\ref{fig:figS1-4}. All three simulations use the same geometry and designed current distribution (a simple homogeneous distribution), differing only in the choice of \( L_c \). For relatively large \( L_c \), a blurring effect occurs, causing a smoothing of the calculated \( J_c(x) \). As \( L_c \) decreases, the sharpness of the reconstructed \( J_c(x) \) significantly improves.

\begin{figure}[!tbh]
\begin{center}
\includegraphics[width=\textwidth]{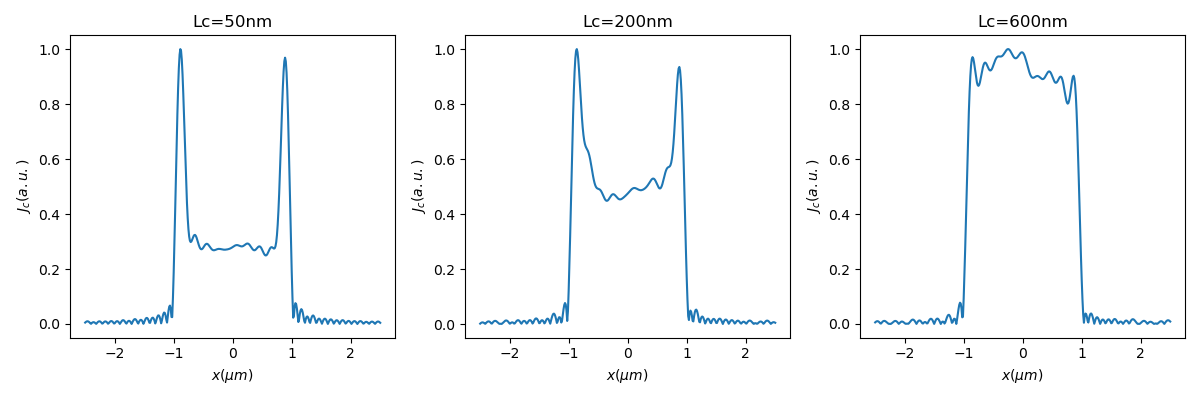}  
\end{center}
\caption{{\bf $|$} Impact of decay length on edge current resolution.} 
\label{fig:figS1-5}
\end{figure} 

This effect is even more pronounced for edge currents. As shown in Fig.~\ref{fig:figS1-5}, when \( L_c \) is relatively large, the edge current can be almost entirely suppressed (panel a). Therefore, accurately estimating the edge current (as described in the previous section) is essential before extracting it from experimental data.

Intuitively, the decay length \( L_c \) acts similarly to a coherence length, facilitating interference across the sample width and effectively averaging out the current density distribution.

\newpage
\section{Tight-Binding Model Simulation}

To visualize the hinge states in \(\text{Bi}_{1-x}\text{Sb}_x\) and compare them to experimental results, we simulate \(\text{Bi}_{1-x}\text{Sb}_x\) using the 16-band tight-binding model~\cite{Liu1995} on a lattice. We first reproduce the eigenvalues at the high-symmetry points (HSPs) for Bi and Sb, as presented in Ref.~\cite{Liu1995}, by solving the continuous tight-binding model. Next, we construct the crystal lattice using Kwant~\cite{Groth2014} with the same parameters. To validate our approach, we successfully reproduce the band structure in the $[1\bar{1}0]$ direction of a bilayer bismuth system, as reported by Shuichi Murakami~\cite{Murakami2006}, as shown in Fig.~\ref{fig:figS1-6}a.

\begin{figure}[!tbh]
\begin{center}
\includegraphics[width=\textwidth]{Figures/Figure_S1-6.png}  
\end{center}
\caption{{\bf $|$ a}, Band structure of a bismuth bilayer with translational symmetry in the $[1\bar{1}0]$ direction. {\bf{b}}, visualization of the hinge modes in the $[1\bar{1}0]$ direction in a 3D slab confined by top and side surfaces with their normals in the $[111]$ and $[001]$ direction respectively. Site color and size indicate the summed density of the modes at the Fermi surface, $\rho_\text{tot}$.} 
\label{fig:figS1-6}
\end{figure} 

Using Kwant, we perform calculations for two types of models: 
1) a finite-size 3D model,
2) an infinite-size model. 
In both cases, we confirm the existence of robust hinge modes across various facet configurations. The model parameters for \(\text{Bi}_{1-x}\text{Sb}_x\) are interpolated as $a = a_{\text{Bi}}(1-x) + a_{\text{Sb}} x$, $x=0.03$.
A direct way to visualize hinge modes is by computing the wavefunctions of the topological states at the Fermi surface and plotting their spatial distribution within the crystal lattice. In our exfoliated flakes, the natural cleaving facet is always (111), occurring between two bilayers due to the stronger intralayer bonding within a bilayer compared to interlayer bonding. We refer to this stacking configuration as AB-AB, where one layer consists of a single AB bilayer.
In Fig.~\ref{fig:figS1-6}b, the hinge states are shown for an eight-bilayer lattice. The hinges are evident and decay both along the $(001)$ plane away from the corners and into the bulk.
As seen in the simulation, the summed wavefunction remains unchanged along the hinge mode direction. This behavior arises because, in the 3D model, adding a lead imposes translational symmetry along that $[1\bar{1}0]$ direction. To explore larger cross-sections while reducing computational costs, we extend the model to the infinite system discussed in the main text.
In an infinite model, it suffices to analyze the cross-section perpendicular to the translational symmetry direction. 



\clearpage
\section{Step-by-step scheme to estimate $I_{c,\ edge}$ and $I_{c,\ bulk}$}
\label{sect:sec2} 

\begin{figure}[!tbh]
\begin{center}
\includegraphics[width=\textwidth]{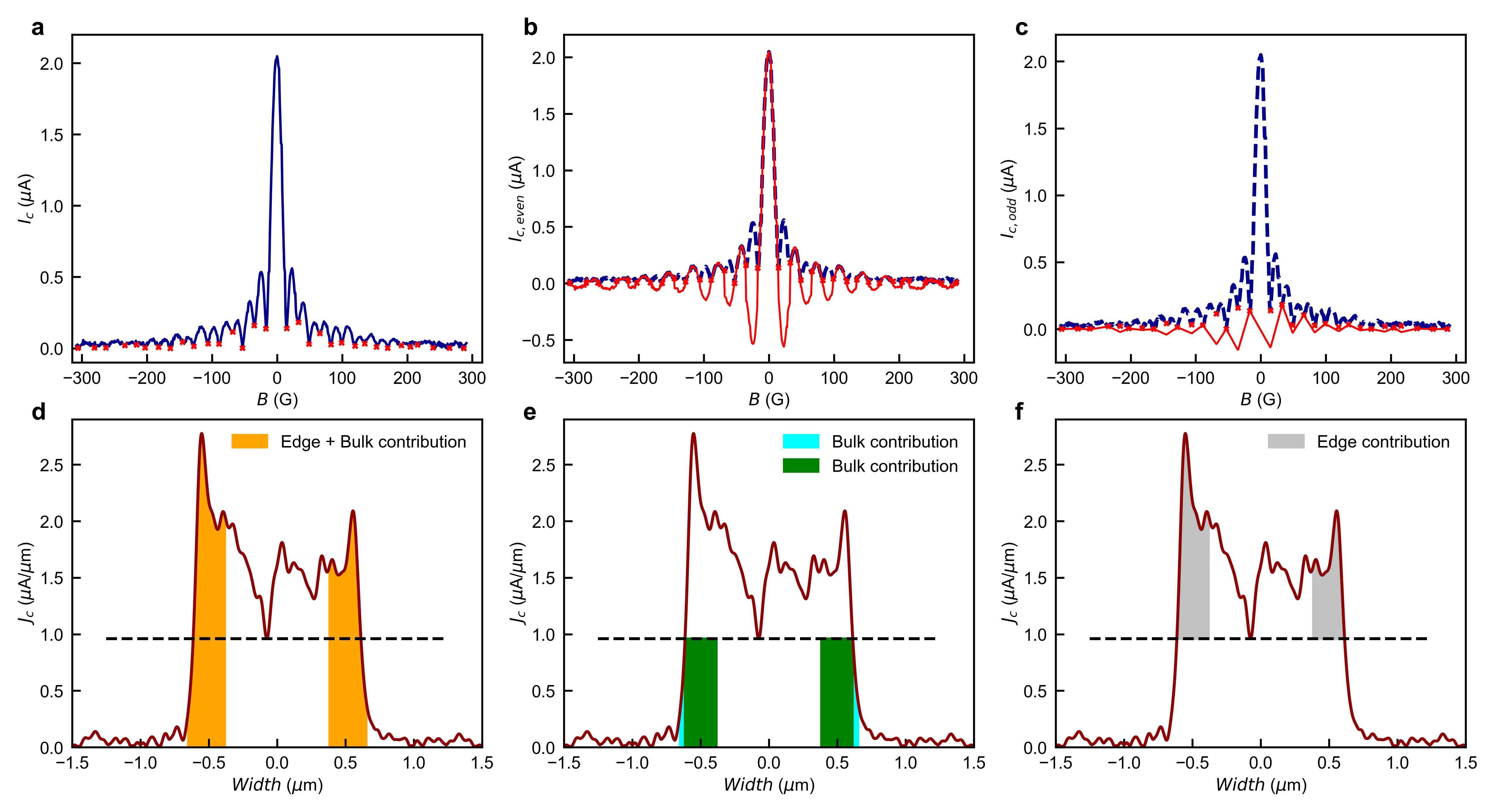}  
\end{center}
\caption 
{{\bf $|$} {\bf a}, An experimentally measured $I_cB$ with red colored cross marking the position of the nodes. {\bf b}, The even component (red solid curve) is calculated by flipping the sign of every second lobe of $I_c$ (darkblue dashed curve). {\bf c}, The odd component (red solid curve) is calculated by interpolating between the alternating nodes of $I_c$ (dark blue dashed curve) but flipping sign of the intermediate node to ensure asymmetry across $B = 0$. {\bf d}, $J_c$ distribution as a function of width of the junction showing peaks at the edges due to presence of hinge modes. Orange shaded region highlights the area under $J_c(x)$ curve, spanning only the edge states width as estimated from the SQI model described in Supplementary Section~\ref{sect:sec1}. This area provides the contribution of both edge states and background bulk, $I_{c,\ (edge\ +\ bulk)}$, to the peak at junction edges. Dashed black line is the minimum value that $J_c(x)$ takes within the physical width of the junction. {\bf e}, $J_c(x)$ profile highlighting only the bulk contribution $I_{c,\ bulk}$ to the peak at junction edges as the area under $J_c(x)$ but below the dashed black line. {\bf f}, $J_c(x)$ profile highlighting only the edge states contribution (gray shaded region), $I_{c,\ edge} = I_{c,\ (edge\ +\ bulk)} - I_{c,\ bulk}$, to the peak at junction edges as the area under $J_c(x)$ but above the dashed black line.} 
\label{fig:figS2}
\end{figure} 

As mentioned in the main text, in a JJ with out-of-plane perpendicular $B$, the $I_cB$ modulation depends strongly on the $J_c(x)$ distribution in the junction expressed by the inverse Fourier transform equation:
\begin{equation}
I_c(B) = \left|\int_{-\infty}^\infty J_c(x) \exp\left(i \frac{2 \pi L_\text{eff} B x}{\Phi_0}\right) \, dx\right|
\label{eq:1}
\end{equation}
where symbols have same meaning as described in main text. Dynes and Fulton \cite{Dynes1971} introduced Fourier techniques to retrieve the $J_c(x)$ from the experimentally measured $I_cB$. The overall supercurrent is composed of both the even and odd parts, i.e. symmetric (cosine) and asymmetric (sine) components.
\begin{equation}
I_{c,\ even}(B) = \int_{-\infty}^\infty J_{c,\ even}(x)\ cos(\frac{2 \pi L_\text{eff} B x}{\Phi_0}) \, dx\ ;\ I_{c,\ odd}(B) = \int_{-\infty}^\infty J_{c,\ odd}(x)\ sin(\frac{2 \pi L_\text{eff} B x}{\Phi_0}) \, dx
\label{eq:2}
\end{equation}
For a symmetric $I_cB$ distribution, the odd component of Eq.~\eqref{eq:1} is zero such that total supercurrent can be expressed with only even component $I_{c,\ even}(B)$. Whereas in presence of an odd component $I_{c,\ odd}(B)$, the total $I_c$ can be expressed with a complex equation:
\begin{equation}
I_c(B) = I_{c,\ even}(B) + iI_{c,\ odd}(B) 
\label{eq:3}
\end{equation}
The Fourier transform of the complex $I_c(B)$ from Eq.~\eqref{eq:3} provides the $J_c(x)$ distribution in the junction expressed as:
\begin{equation}
J_c(x) = \alpha \left|\int_{-B}^B I_c(B) \exp\left(-i \frac{2 \pi L_\text{eff} B x}{\Phi_0}\right) \, dB\right|
\label{eq:4}
\end{equation}
where $\alpha$ is the scaling factor calculated using the experimentally measured maximum $I_c$ value at $B = 0$.

\newpage
\section{Supercurrent characteristics of F2\_800 nm Josephson junction}
\label{sect:sec3} 

\begin{figure}[!tbh]
\begin{center}
\includegraphics[width=\textwidth]{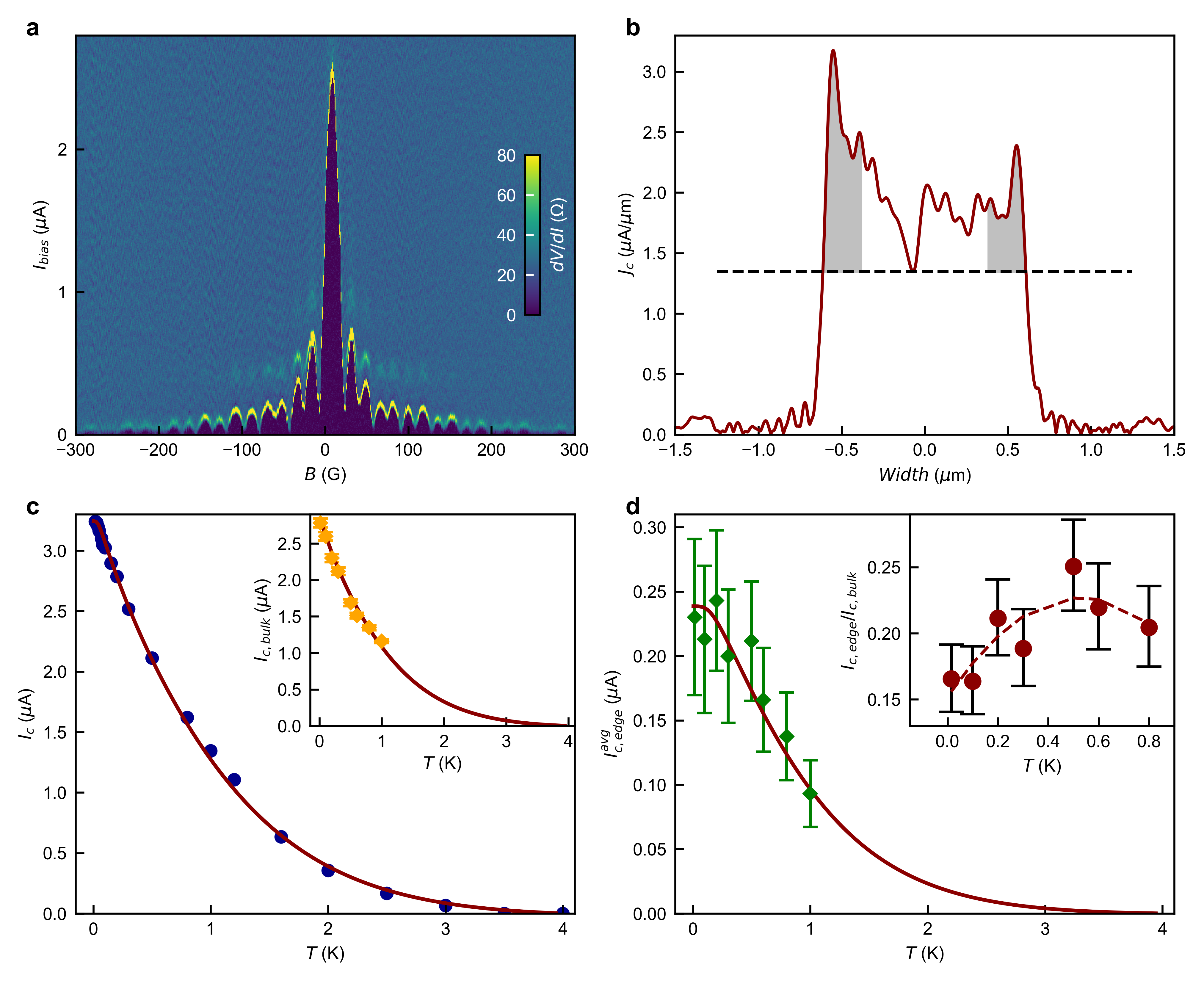}  
\end{center}
\caption 
{{\bf $|$} {\bf a}, $I_cB$ map at 300 mK for 800 nm junction. The colormap represents the differential resistance dV/dI of the device with $R_N = 25\ \Omega$. {\bf b}, Extracted $J_c(x)$ profile across the junction width showing enhanced density at edges typical for a junction featuring edge states. Edge states width of 266 nm has been estimated from our simulation. Gray shaded area highlights the edge-propagated $I_c$, which we estimate $I_{c,\ left\ edge} = 267.66$ nA and $I_{c,\ right\ edge} = 131.86$ nA. Dashed black line represents the minimum value of the bulk background contribution. {\bf c}, Temperature dependence of total (both edges + bulk) $I_c$ for this device. Experimental $I_cT$ (darkblue dots) can be explained clearly with the Eilenberger theory (solid dark red). The estimated fitting parameters are $D = 0.9998$, $\xi = 260$ nm, $T_c = 4$ K and $R_N = 227\ \Omega$. Inset shows the temperature dependence of extracted $I_{c,\ bulk}$ from ({\bf b}) which has been fitted with Eilenberger theory with fit parameters $D = 0.9998$, $\xi = 260$ nm, $T_c = 4$ K and $R_N = 266.8\ \Omega$. Error bars originate from the standard deviation between experimental and numerically calculated $J_c(x)$ pattern. {\bf d}, Extracted average $I^{avg}_{c,\ edge}$ given by $(I_{c,\ left\ edge} + I_{c,\ right\ edge})/2$ variation with temperature (green diamonds). The Eilenberger theory fit parameters are $D = 0.99$, $\xi = 200$ nm, $T_c = 4$ K and $R_N = 2070.8\ \Omega$. Average number of channels per edge are 6 modes. Inset shows the $I_{c,\ edge}/I_{c,\ bulk}$ ratio ($I_{c,\ edge} = I_{c,\ left\ edge} + I_{c,\ right\ edge}$) for the 800 nm junction showing a similar increasing trend with temperature as seen for device F2\_600 nm (Fig. 3d, main text). The dashed curve is guide-to-the-eye.} 
\label{fig:figS3}
\end{figure}

\newpage
\section{Supercurrent characteristics of F2\_1000 nm Josephson junction}
\label{sect:sec4} 

\begin{figure}[!tbh]
\begin{center}
\includegraphics[width=\textwidth]{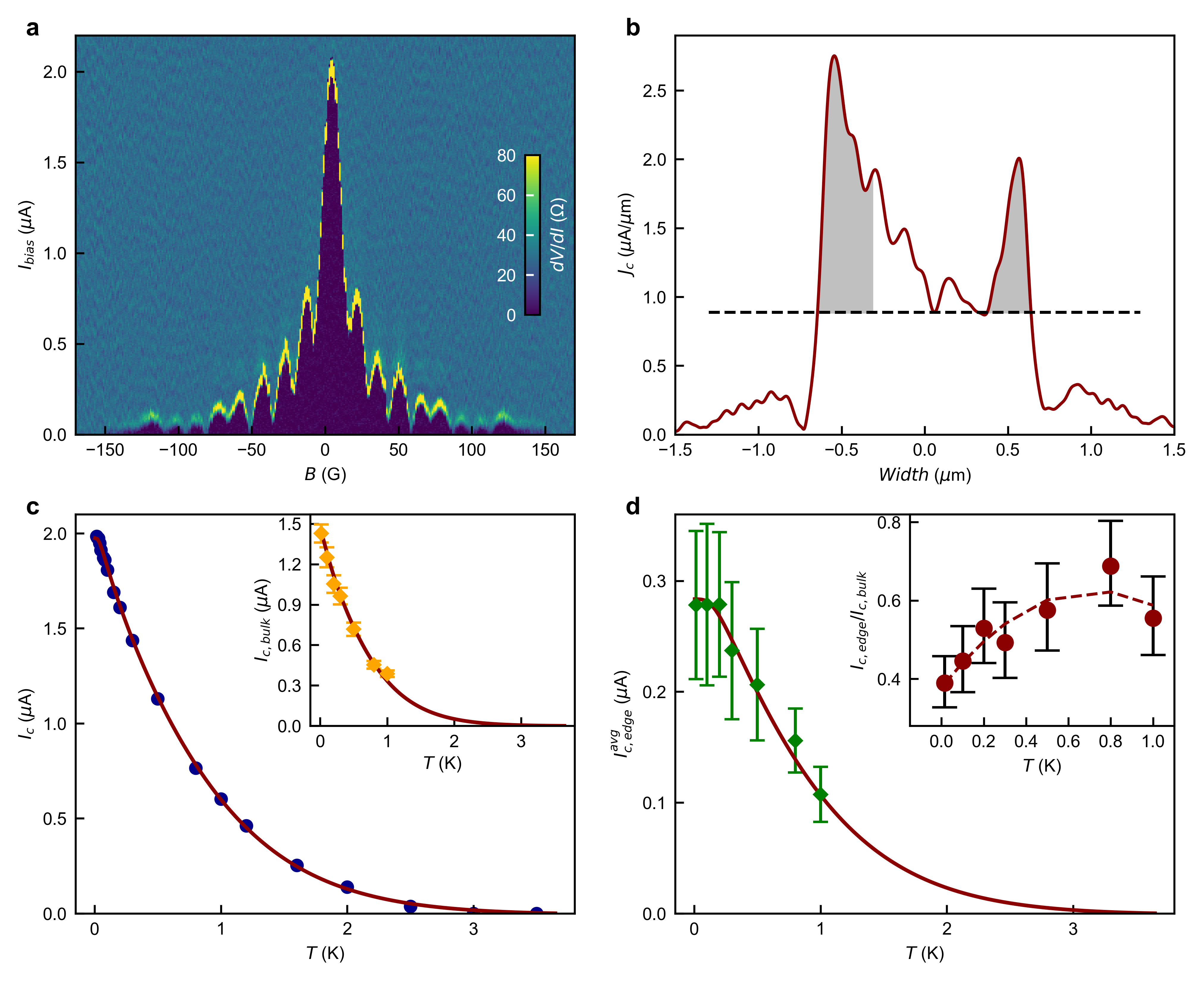}  
\end{center}
\caption 
{{\bf $|$} {\bf a}, $I_cB$ map at 15 mK for 1000 nm junction. The colormap represents the differential resistance dV/dI of the device. $R_N$ is 30 $\Omega$ for this device. {\bf b}, Extracted $J_c(x)$ profile across the junction width showing enhanced density at edges typical for a junction featuring edge states. Edge states width of 381 nm has been estimated from our simulation. Gray shaded area highlights the edge-propagated $I_c$, which we estimate to be $I_{c,\ left\ edge} = 403.41$ nA and $I_{c,\ right\ edge} = 152.83$ nA. Dashed black line represents the minimum value of the bulk background contribution. {\bf c}, Temperature dependence of total (both edges + bulk) $I_c$ for this device. Experimental $I_cT$ (dark blue dots) can be explained clearly with the Eilenberger theory (solid dark red). The estimated fitting parameters are $D = 0.9998$, $\xi = 260$ nm, $T_c = 3.7$ K and $R_N = 302.6\ \Omega$. Inset shows the temperature dependence of extracted $I_{c,\ bulk}$ from ({\bf b}) which has been fitted with Eilenberger theory with fit parameters $D = 0.9998$, $\xi = 200$ nm, $T_c = 3.7$ K and $R_N = 354.4\ \Omega$. Error bars originate from the standard deviation between experimental and numerically calculated $I_cB$ pattern. {\bf d}, Extracted average $I^{avg}_{c,\ edge}$ given by $(I_{c,\ left\ edge} + I_{c,\ right\ edge})/2$ variation with temperature (green diamonds). The Eilenberger theory fit parameters are $D = 0.99$, $\xi = 260$ nm, $T_c = 3.7$ K and $R_N = 1652.6\ \Omega$. Average number of channels per edge are 8 modes. Inset shows the $I_{c,\ edge}/I_{c,\ bulk}$ ratio ($I_{c,\ edge} = I_{c,\ left\ edge} + I_{c,\ right\ edge}$) for the 1000 nm junction showing a similar increasing trend with temperature as seen for the other two junctions on this flake. The dashed curve is guide-to-the-eye.} 
\label{fig:figS4}
\end{figure}

\newpage
\section{RF dependence of missing odd Shapiro steps in F2\_600 nm junction}
\label{sect:sec5}

\begin{figure}[!tbh]
\begin{center}
\includegraphics[width=\textwidth]{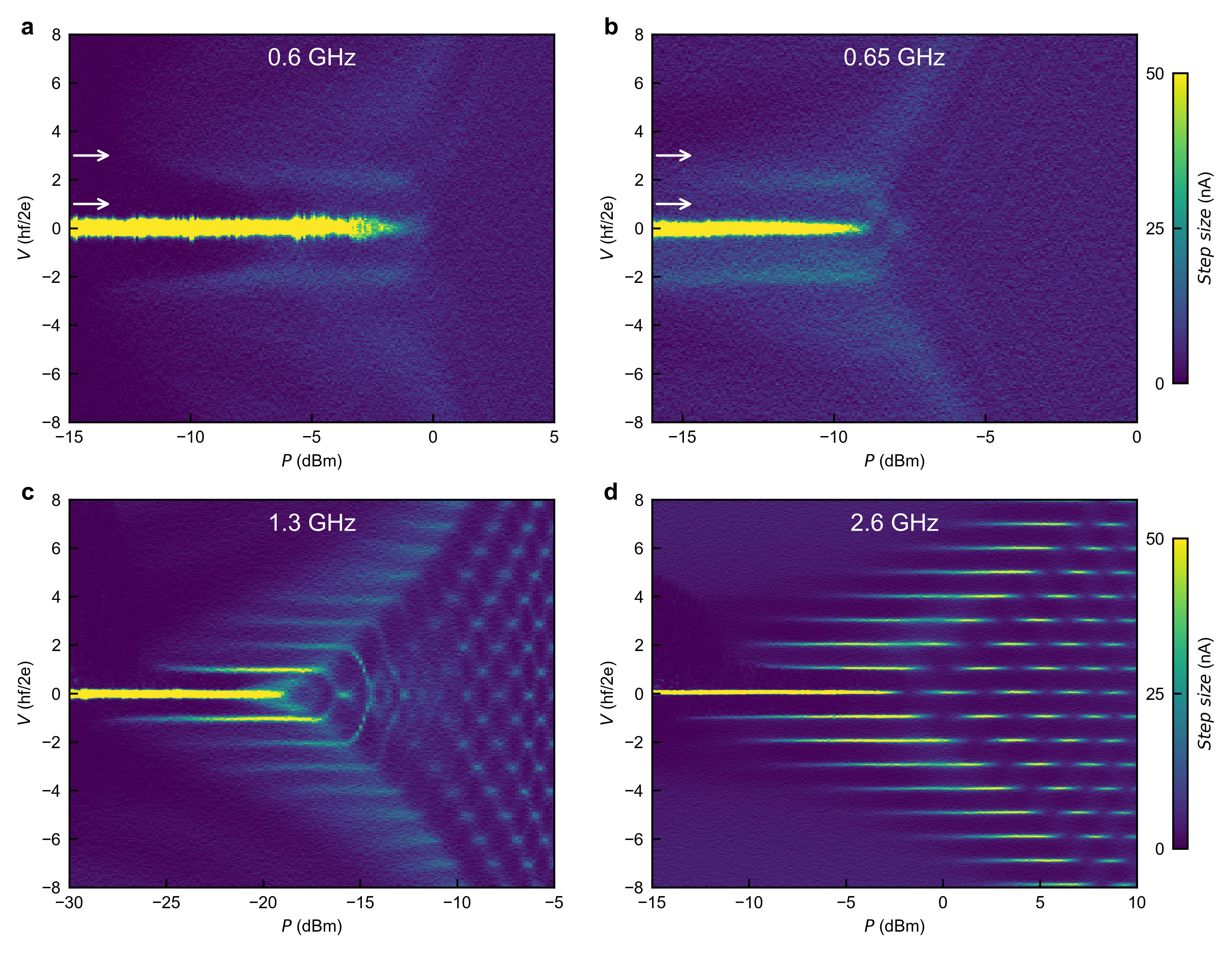}  
\end{center}
\caption 
{{\bf $|$} Re-appearance of the odd Shapiro steps at higher frequency for device F2\_600 nm. White arrows mark the positions of the missing first and third odd integer steps at $f =$ ({\bf a}) $0.6$ GHz and ({\bf b}) $0.65$ GHz. Figure 3a of main text shows the missing steps at $0.9$ GHz. ({\bf a}) and ({\bf b}) share the same voltage bin size and the colorscale. All the Shapiro steps are present at higher RF, $f = $ ({\bf c}) $1.3$ GHz and ({\bf d}) $2.6$ GHz. ({\bf c}) and ({\bf d}) share the same voltage bin size and the colorscale.} 
\label{fig:figS5}
\end{figure}

\newpage
\section{Temperature dependence of missing odd Shapiro steps in F2\_600 nm junction}
\label{sect:sec6}

\begin{figure}[!tbh]
\begin{center}
\includegraphics[width=\textwidth]{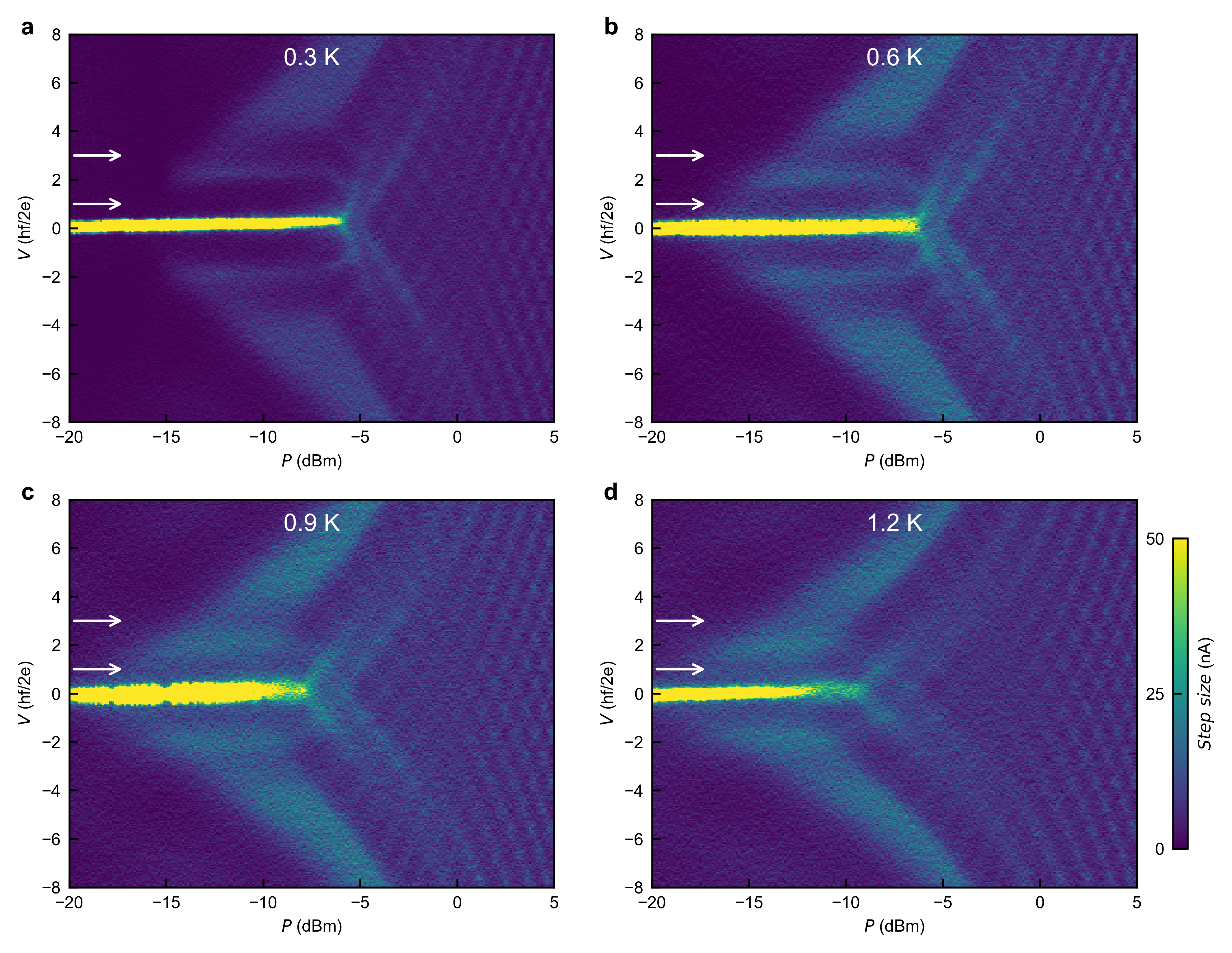}  
\end{center}
\caption 
{{\bf $|$} Robustness of the missing Shapiro steps with increasing temperature for device F2\_600 nm. Same voltage bin size has been used for all the temperatures. White arrows mark the positions of the missing first and third odd integer steps at $T =$ ({\bf a}) $0.3$ K, ({\bf b}) $0.6$ K, ({\bf c}) $0.9$ K and ({\bf d}) $1.2$ K. Thermal smearing of the quantized voltage steps can be seen with increasing temperature.} 
\label{fig:figS6}
\end{figure}



\newpage
\section{Missing odd Shapiro steps in F2\_800 nm junction}
\label{sect:sec7}

\begin{figure}[!tbh]
\begin{center}
\includegraphics[width=0.8\textwidth]{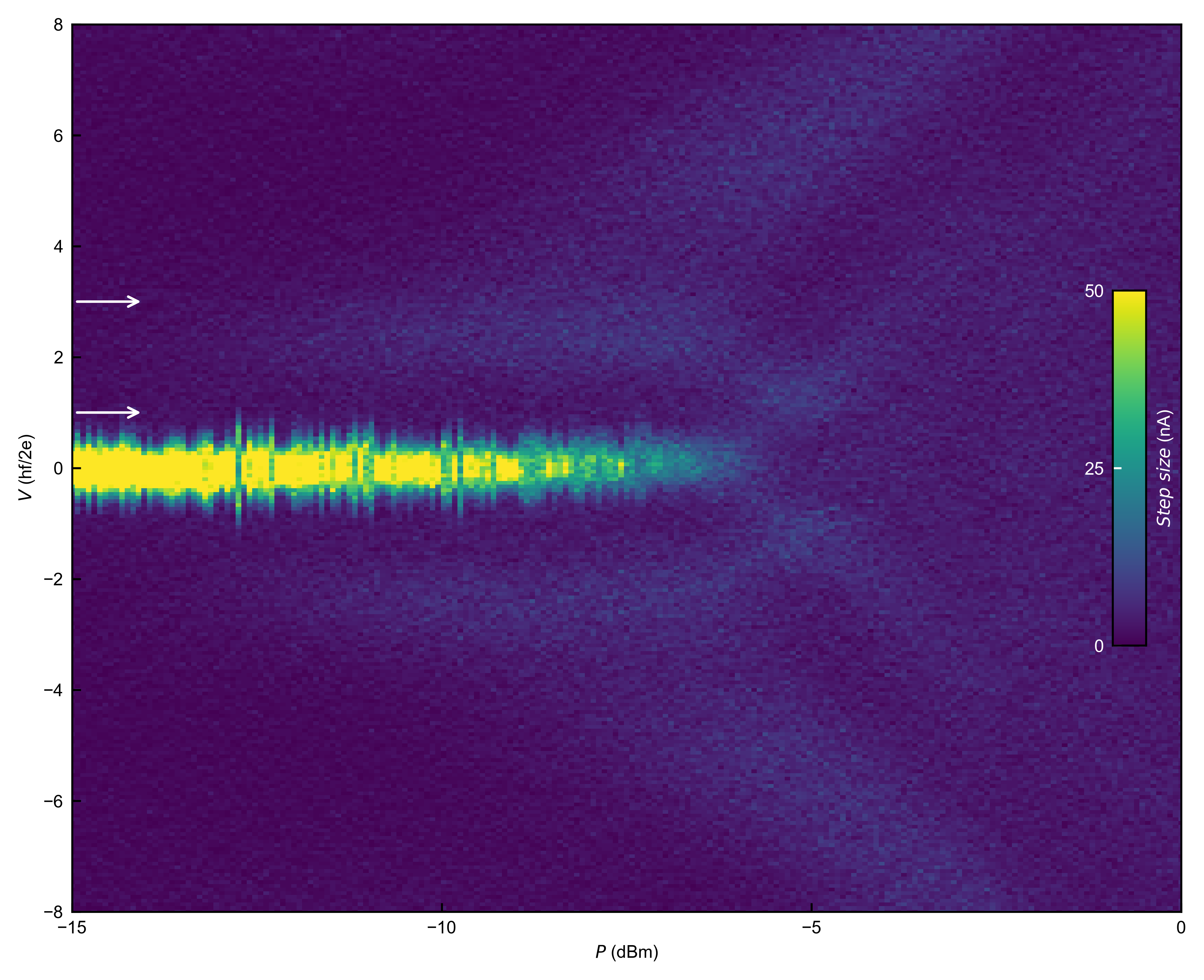}  
\end{center}
\caption 
{{\bf $|$} {\bf a}, Missing odd steps ($n = 1,\ 3$) for the F2\_800 nm device under excitation with low RF irradiation $f = 0.65$ GHz at 70 mK. White arrows mark the positions of the missing Shapiro steps.} 
\label{fig:figS7}
\end{figure}

\newpage
\section{Supercurrent characteristics of F1 bulk-edge Josephson junction device}
\label{sect:sec8}

\begin{figure}[!tbh]
\begin{center}
\includegraphics[width=\textwidth]{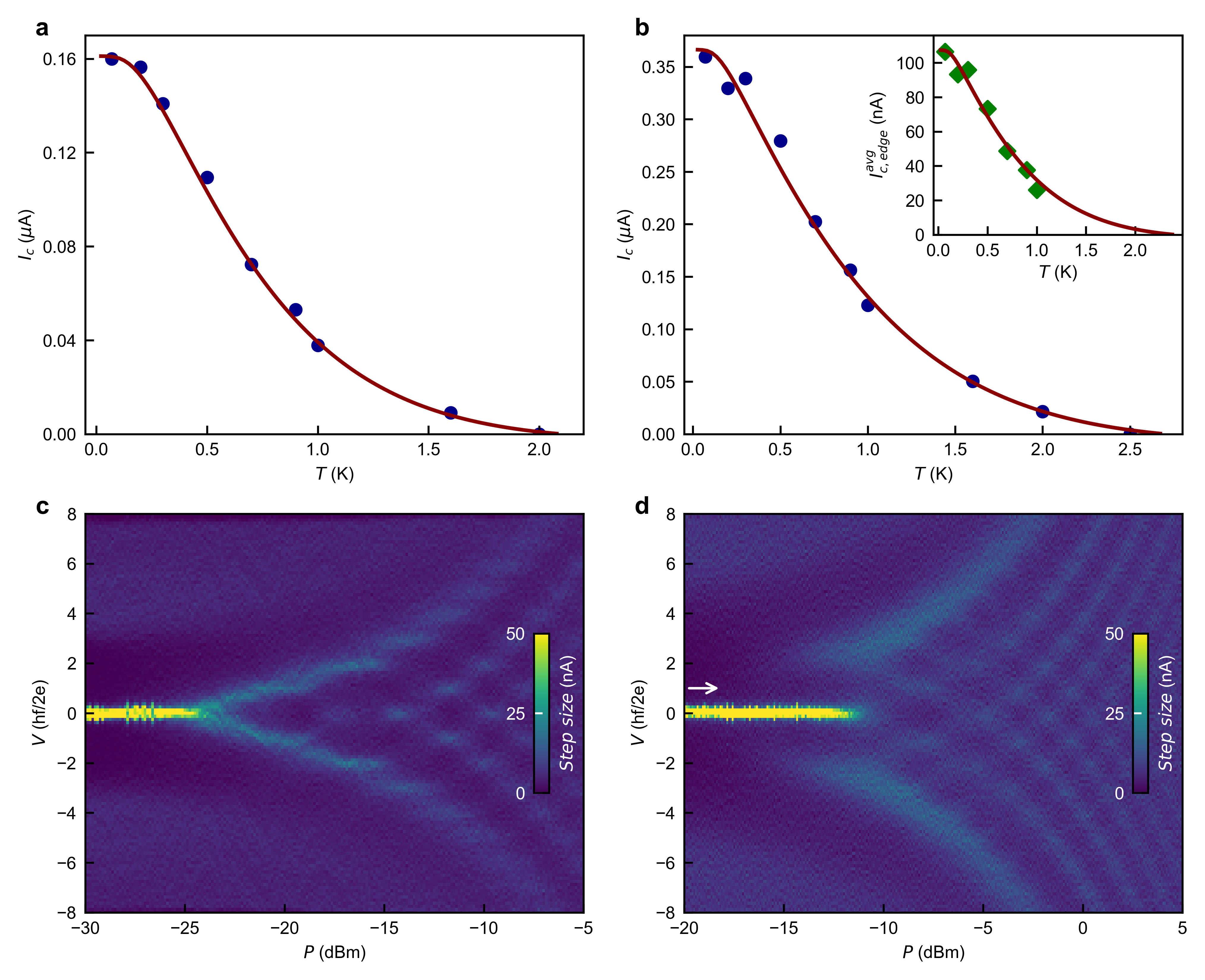}  
\end{center}
\caption 
{{\bf $|$} {\bf a}, Temperature dependence of total $I_c$ for device F1\_bulk JJ with $L = 800$ nm. Experimental data points (dark blue) can be fitted with Eilenberger theory (solid dark red). The estimated fitting parameters are $D = 0.95$, $\xi = 300$ nm, $T_c = 2.1$ K and $R_N = 1530\ \Omega$. {\bf b}, Temperature dependence of total $I_c$ for device F1\_edge JJ with $L = 800$ nm can also be fitted with Eilenberger theory. The estimated fitting parameters are $D = 0.99$, $\xi = 340$ nm, $T_c = 2.7$ K and $R_N = 1234.5\ \Omega$. Inset shows the temperature dependence of the extracted $I_c$ for the edge modes (green diamonds) given by $(I_{c,\ left\ edge} + I_{c,\ right\ edge})/2$. The Eilenberger theory fit parameters are $D = 0.99$, $\xi = 340$ nm, $T_c = 2.4$ K and $R_N = 3743.3\ \Omega$. {\bf c, d}, RF measurements showing the Shapiro steps in F1\_bulk-edge JJ devices at 70 mK. While all the integer steps are observed for the bulk JJ ({\bf c}) at 0.85 GHz, there is a missing odd $n = 1$ step for the edge JJ ({\bf d}) at 0.8 GHz marked by white arrow. We did not observe missing odd steps for the bulk JJ at any other frequency. This is a strong evidence of topological hinge modes in this material.} 
\label{fig:figS8}
\end{figure}

\newpage
\section{Monotonically decaying $I_cB$ for different bulk Josephson junctions}
\label{sect:sec9}

\begin{figure}[!tbh]
\begin{center}
\includegraphics[width=\textwidth]{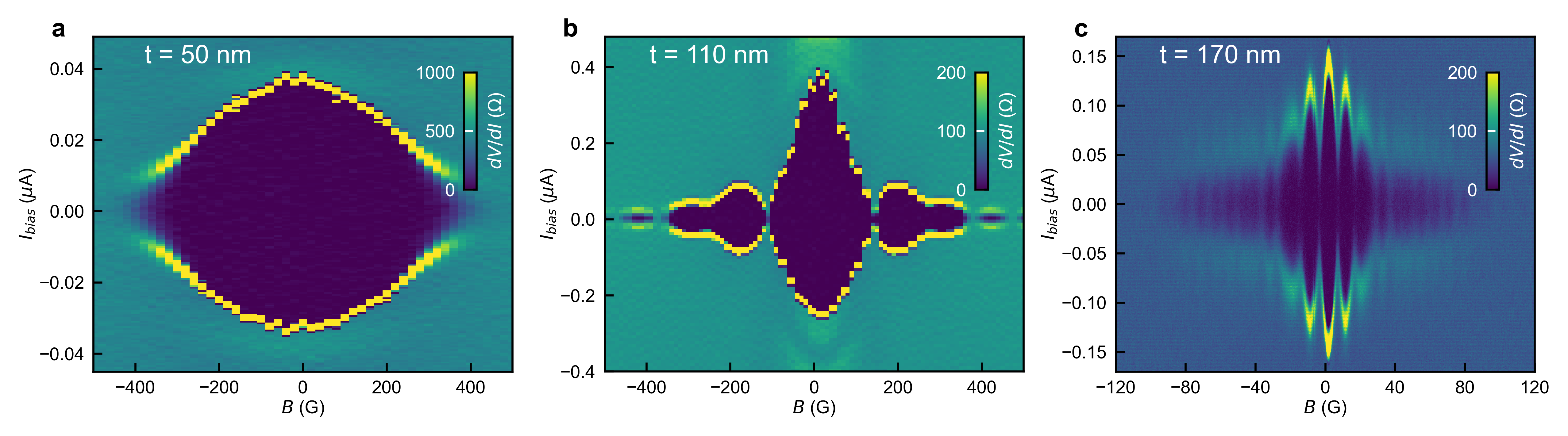}  
\end{center}
\caption 
{{\bf $|$} Monotonically decaying $I_cB$ pattern is observed for three different bulk JJs on separate flakes (curves overlayed for clarity). Dimensions of the bulk JJ device are as follows: $L=600$ nm, $W=1.2\ \mu$m, $t=150$ nm for green curve, $L=800$ nm, $W=1.2\ \mu$m, $t=150$ nm for blue curve and $L=800$ nm, $W=1\ \mu$m, $t=200$ nm for red curve.}
\label{fig:figS9}
\end{figure}

\newpage
\section{Surface morphology of the flake F2 probed with atomic force microscopy}
\label{sect:sec10}

\begin{figure}[!tbh]
\begin{center}
\includegraphics[width=0.6\textwidth]{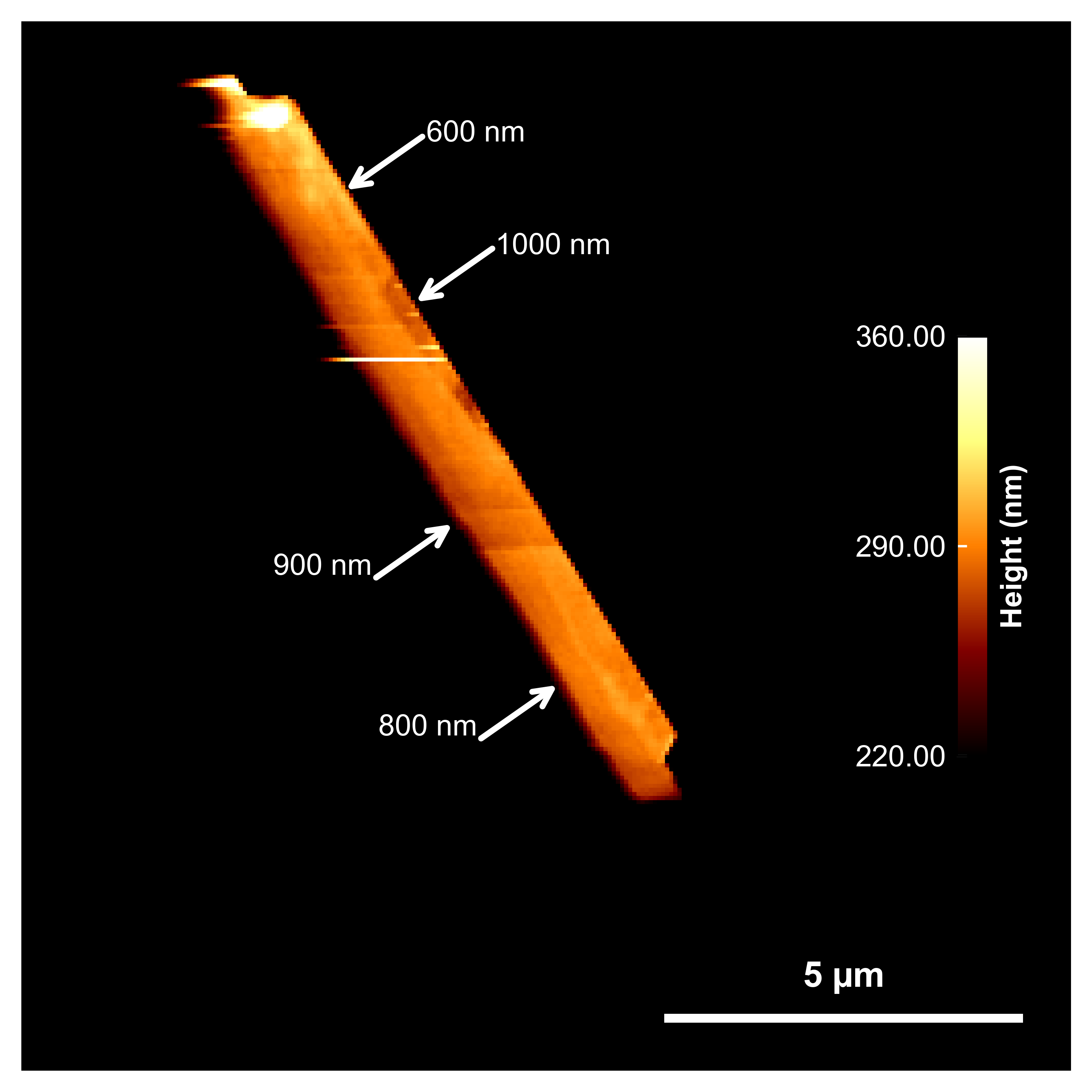}  
\end{center}
\caption 
{{\bf $|$} Atomic force microscopy (AFM) image of flake F2 highlighting the surface morphology and positions of the different junctions marked by white arrows. The thickness ($\sim250$ nm) of the flake is uniform throughout it's entire length. The colorscale is chosen in a way that it increases the contrast and provides better resolved surface morphology.} 
\label{fig:figS10}
\end{figure} 

Figure \ref{fig:figS10} shows the AFM height sensor image for flake F2 with position of the four different junctions fabricated on this flake marked by white arrows (see Fig. 2a for device image). An additional large step ($\sim15$ nm high) on one of the edges can be seen for the 1000 nm junction. Interestingly, in order to emulate the $I_cB$ and $J_c(x)$ of this junction, we had to consider this step on one edge in our simulation too. A small bump can be seen at the center of flake F2 where 800 nm junction is placed, which also renders a small bump in its $J_c(x)$ distribution around the middle of the junction width (Fig. \ref{fig:figS3}b). This shows the sensitivity of Dynes and Fulton Fourier techniques to the real surface morphology of the junction.


\newpage
\section{SEM characterization}
\label{sect:sec11}

\begin{figure}[!tbh]
\begin{center}
\includegraphics[width=\textwidth]{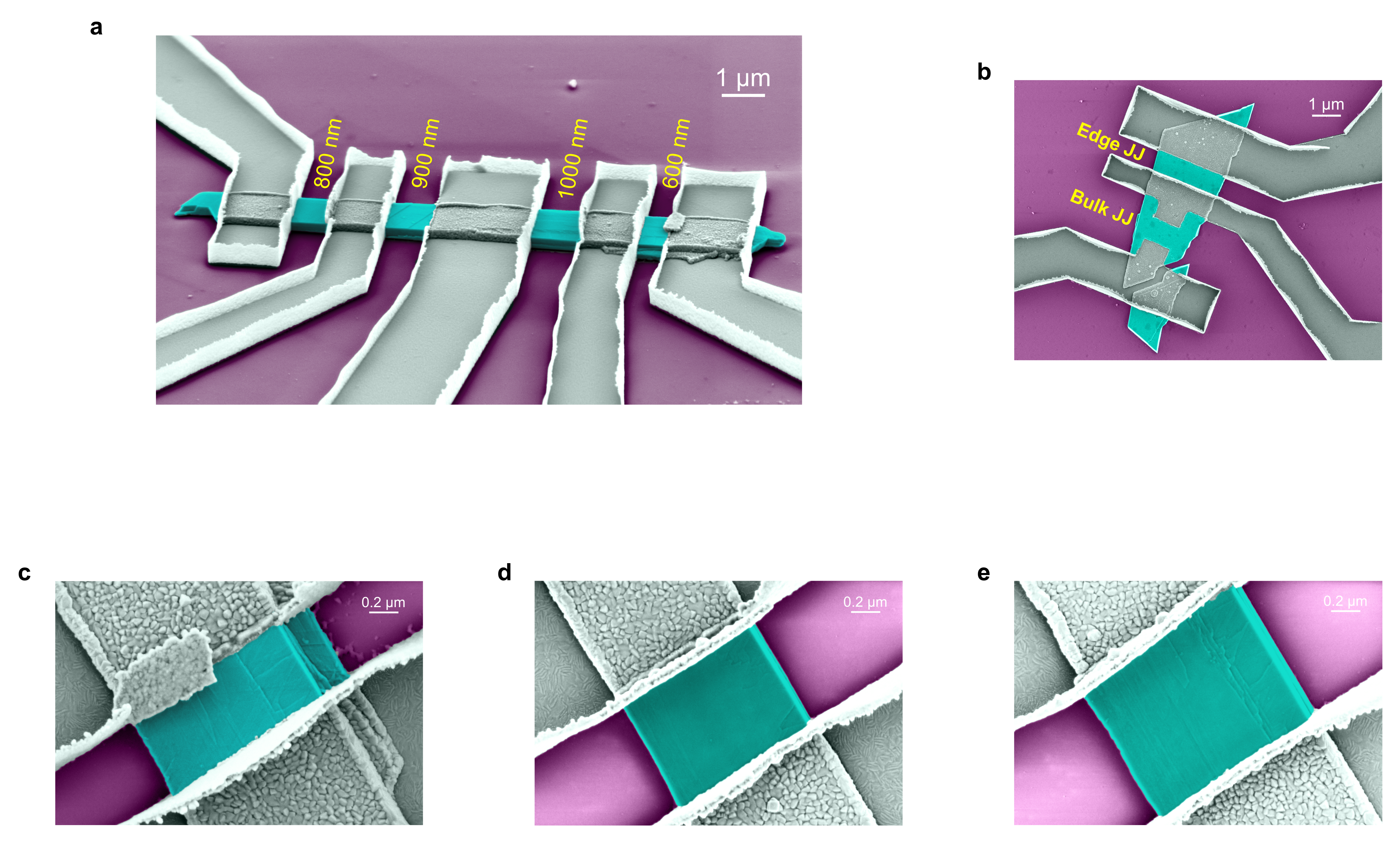}  
\end{center}
\caption 
{{\bf $|$} {\bf a}, Cross-sectional SEM image of flake F2 highlighting the different junctions. {\bf b}, SEM image of flake F1 showing the bulk and edge JJs both 800 nm long. {\bf c-e}, SEM images of $600$, $800$ and $1000$ nm junctions in \textbf{a}.} 
\label{fig:figS11}
\end{figure}

\newpage
\section{$Q_{12}$ calculation in RF Shapiro steps measurements}
\label{sect:sec12}

\begin{figure}[!tbh]
\begin{center}
\includegraphics[width=\textwidth]{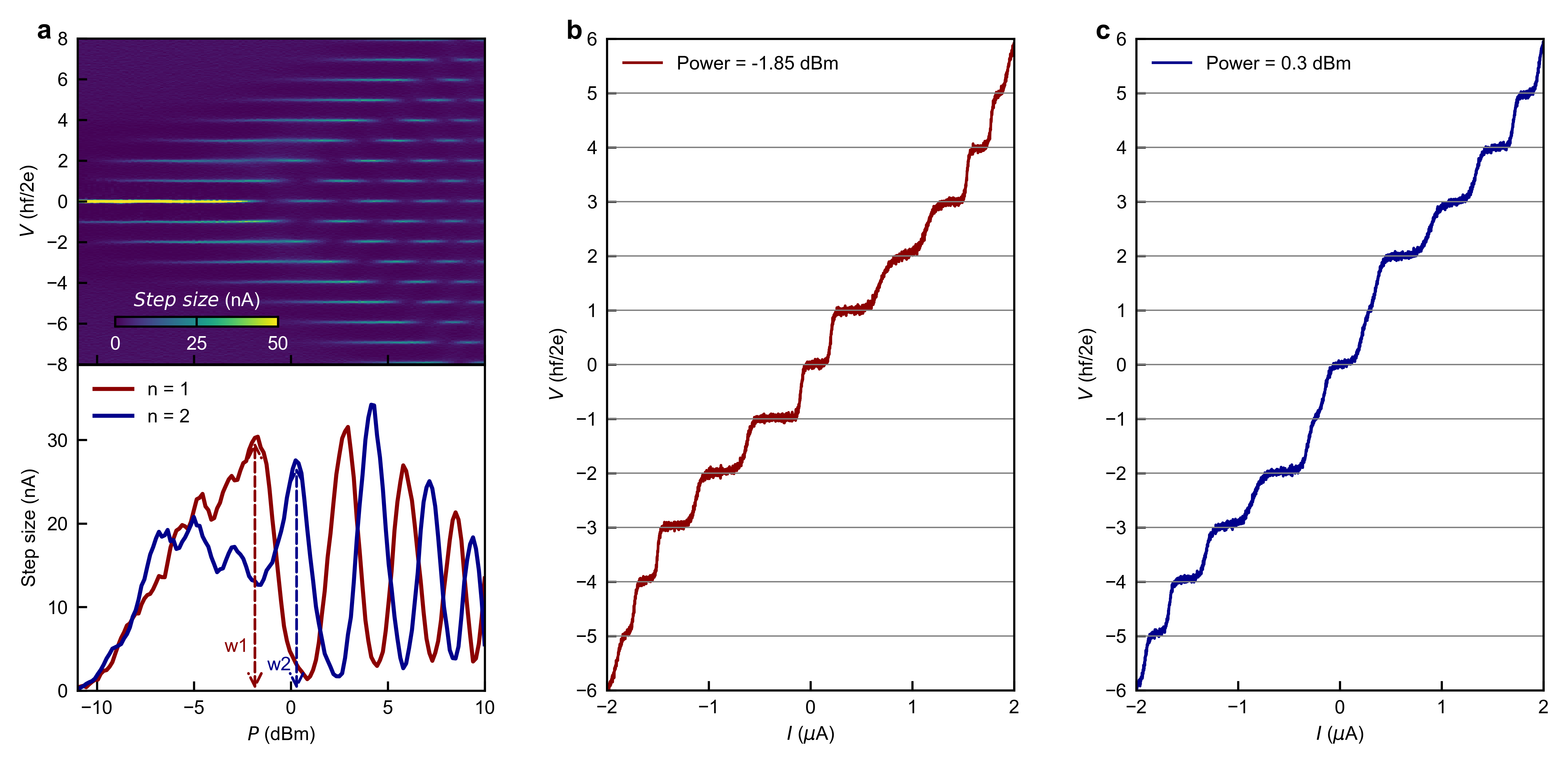}  
\end{center}
\caption 
{{\bf $|$} {\bf a}, Colormap showing the Shapiro steps for device F2$\_600$ nm under 2.65 GHz RF irradiation at 70mK. Bottom panel are the line-cuts from top panel at $n=1$ and $2$ depicting the power dependence of their step sizes. $w1$ and $w2$ mark the step size width of $n=1$ and $2$ as the maximum value of first lobe in the power dependence of step size. The ratio $Q_{12}=w1/w2$, thus provides information about the suppression of $n=1$ step. {\bf b,c}, $IV$ curves (raw data) with voltage normalized to $\frac{hf}{2e}$ at powers of -1.85 dBm ($w1$) and 0.3 dBm ($w2$) showing the quantized step sizes for reference.} 
\label{fig:figS12}
\end{figure}

\newpage
\section{Supercurrent characteristics of a shorter ($L=350$ nm) Josephson junction}
\label{sect:sec13}

\begin{figure}[!tbh]
\begin{center}
\includegraphics[width=\textwidth]{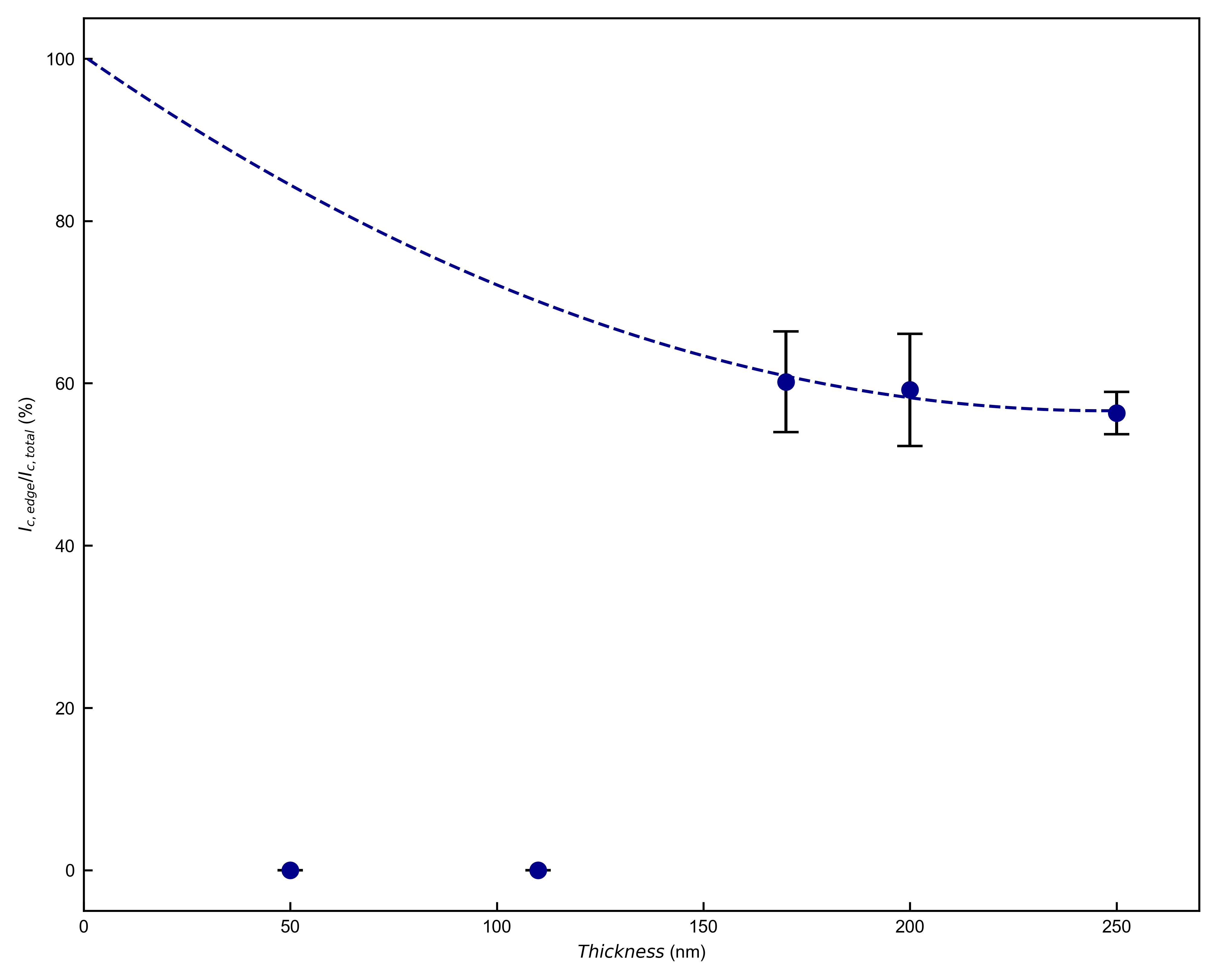}  
\end{center}
\caption 
{{\bf $|$} {\bf a}, Shorter junctions allow both bulk and edge channels to survive ballistically, producing the standard single-slit Fraunhofer interference. {\bf b}, Extracted $J_c(x)$ distribution of across this junction width shows bulk-dominated conduction. In contrast, the bulk coherence decays for longer junctions, leaving only robust edge channels that dominate and yield SQUID-like patterns as shown in previous devices. This transition reflects a length-dependent crossover between the bulk and edge transport channels.} 
\label{fig:figS13}
\end{figure}

\newpage
\section{Table: Summary of device parameters and properties}
\label{sect:sec11}
\begin{table}[h]
    \centering
    \begin{tabularx}{\textwidth}{|X|X|X|X|X|X|X|}  
        \hline
        \textbf{$Device$ \#} & \textbf{$t\ (nm)$} & \textbf{$L\ (nm)$} & \textbf{$W\ (\mu m)$} & \textbf{$I_cR_N\ (\mu V)$} & \textbf{$I_cB$} & \textbf{$RF$} \\
        \hline
        BS04\_1 & 50 & 300 & 0.3 & 16 & Monotonic decay & Not measured \\
        \hline
        BS04\_2 & 110 & 300 & 0.55 & 40 & Typical Fraunhofer & Not measured \\
        \hline
        BS07\_CF1-bulk & 150 & 600 & 1.2 & 4.7 & Monotonic decay & Not measured \\
        \hline
        BS07\_BF8-bulk & 150 & 800 & 1.2 & 14.1 & Monotonic decay & Not measured \\
        \hline
        BS06\_F5 & 170 & 1000 & 1.7 & 7.7 & SQUID-like & All steps present \\
        \hline
        BS06\_F12 & 175 & 350 & 1 & 25.7 & Typical Fraunhofer & All steps present \\
        \hline
        BS07\_F1-edge & 200 & 800 & 2.7 & 9.7 & SQUID-like & Missing $n = 1$ \\
        \hline
        BS07\_F1-bulk & 200 & 800 & 1 & 7.2 & Monotonic decay & All steps present \\
        \hline
        BS07\_F2 & 250 & 600 & 1.4 & 82.5 & SQUID-like & Missing $n = 1,\ 3$ \\
        \hline
        BS07\_F2 & 250 & 800 & 1.3 & 77.5 & SQUID-like & Missing $n = 1,\ 3$ \\
        \hline
        BS07\_F2 & 250 & 900 & 1.4 & 50.4 & SQUID-like & Not measured \\
        \hline
        BS07\_F2 & 250 & 1000 & 1.4 & 56.1 & SQUID-like & All steps present \\
        \hline
    \end{tabularx}
    \vspace{5mm}
    \caption{{\bf $|$} Summary of device parameters and properties. ($R_N$ is from experimental $dV/dI$)}
    \label{tab:tab1}
\end{table}

\bibliographystyle{naturemag}
\bibliography{reference}

\vspace{1ex}
\end{spacing}